\documentclass[12pt,prd,longbibliography,hyperref]{revtex4-1}
\pdfoutput=1
\usepackage{amssymb}
\usepackage{amsmath}
\usepackage{graphicx}
\usepackage{hyperref}
\usepackage{color}
\newcommand{\be}{\begin{equation}}
\newcommand{\ee}{\end{equation}}
\newcommand{\bea}{\begin{eqnarray}}
\newcommand{\eea}{\end{eqnarray}}

\newcommand{\f}[2]{\frac{#1}{#2}}
\newcommand{\grad}{\vec \nabla}
\newcommand{\deltabar}{\ensuremath{\bar {\triangle}}}
\newcommand{\cross}{\times}
\newcommand{\sign}{{\rm sgn}}
\newcommand{\ket}[1]{|#1\rangle}
\newcommand{\bra}[1]{\langle #1|}

\newcommand{\Poincare}{Poincar\'e\ }
\renewcommand{\Im}{{\rm{\,Im\!}}}

\begin{document}
\title{A Gauge Field Theory of Continuous-Spin Particles}
\author{Philip Schuster}
\email{pschuster@perimeterinstitute.ca}
\affiliation{Perimeter Institute for Theoretical Physics,
Ontario, Canada, N2L 2Y5 }
\author{Natalia Toro}
\email{ntoro@perimeterinstitute.ca}
\affiliation{Perimeter Institute for Theoretical Physics,
Ontario, Canada, N2L 2Y5 }
\date{\today}
\begin{abstract}
We propose and quantize a local, covariant gauge-field action  
that unifies the description of all free helicity and continuous-spin degrees of freedom in a simple manner. 
This is the first field-theory action of any kind for continuous spin particles; it is consistent as a quantum theory and generalizes to any number of dimensions.  
The fields live on the null cone of an internal four-vector ``spin-space''; 
in $D$ dimensions a linearized gauge invariance reduces their physical content to a single function on a Euclidean 
$(D-2)$-plane, on which the little group $E(D-2)$ acts naturally.  
A projective version of the action further reduces the physical content to $S^{D-3}$, enabling a new local description of 
particles with any spin structure, and in particular a tower of all integer-helicity particles for $D=4$.   
Gauge-invariant interactions with a background current are added in a straightforward manner. 
\end{abstract}
\maketitle
\tableofcontents
\newpage
%%%%%%%%%%%%%%%%%%%%%%%%%%%%%%%
\section{Introduction}

Two recent results hint that continuous-spin particles (CSPs) can have non-trivial interactions with matter and could potentially be relevant to Nature.  CSPs are the most general massless particles allowed by Poincar\'e-invariance \cite{Wigner:1939cj}. They are characterized by a scale $\rho$, and at $\rho=0$ reduce to familiar integer or half-integer helicity particles.
Along with  particles of spin $\leq 2$, CSPs are the only massless states possessing covariant 
soft factors, which opens the possibility that CSPs may mediate long-range forces \cite{Schuster:2013pxj}. Even more striking is that the soft factors and certain candidate CSP amplitudes approach scalar, photon, and graviton soft factors and amplitudes at energies larger than $\rho$ \cite{Schuster:2}.  

These findings underscore the value of seeking a bulk space-time theory for CSPs, which we present here for the first time.  
Our reasons for doing so are severalfold. First, we can --- a healthy free quantum theory does exist, despite 40 years of lore to the contrary \cite{Yngvason:1970fy,Iverson:1971hq,Chakrabarti:1971rz,Abbott:1976bb,Hirata:1977ss}!  More importantly though, we want to develop a physical interpretation of CSPs that permits the study of classical phenomena and improved computational control over perturbative processes relative to $S$-matrix constructions.  The most pressing question that a field theory should clarify is the origin of the apparently non-local phase in the soft factors of \cite{Schuster:2013pxj}, which are bounded and analytically tame but may hint at an underlying obstruction to a fully consistent interacting theory.
Though our understanding of the new gauge field theory we propose is still rudimentary, the theory has dramatically extended our perspective on the above,
and has sharply defined several avenues for further investigation.   

The free action is 
\be
S_{free}  =  \frac{1}{2}\int d^4 x d^4\eta \left[ \delta'(\eta^2) (\partial_x \psi)^2 + \frac{1}{2}\delta(\eta^2)(\Delta\psi)^2 \right], \label{eq:BaseAction}
\ee
where $\delta'(x)=\frac{d}{dx}\delta(x)$, $\Delta = \partial_\eta.\partial_x + \kappa$, and $\kappa$ is a dimensionful coefficient. 
The field $\psi(\eta,x)$ depends on an internal Minkowski ``spin-space'' four-vector $\eta^\mu$ in addition to the usual position-space coordinate $x^\mu$.  Though we have enlarged the coordinate space of $\psi$, $\eta^\mu$ should not be confused with a new space-time coordinate. There are no kinetic terms $\partial_\eta^2$ and both terms in the action have only two $x$-derivatives.  The global space-time symmetry is the usual $ISO(3,1)$ --- translations act only on $x^\mu$, while homogeneous Lorentz transformations act on both $x$ and $\eta$.  This suggests that the $\eta$-space should be interpreted as an internal space encoding spin, an intuition we will build up shortly.  It is evident that rescaling $\kappa \rightarrow a \kappa $ with $a\ne 0$ and redefining $\eta \rightarrow a^{-1} \eta$ leaves the action unchanged.  Thus the value of $\kappa$ has no physical significance, except for whether it is zero or not. 
When $\kappa\neq 0$, this action describes a family of CSPs with every positive spin-scale $\rho>0$,
while for $\kappa=0$ it describes towers of particles of all integer helicities.
The description is highly redundant --- the action is invariant under
\be
\delta\psi(\eta,x) = \left( \eta \cdot \partial_x - \frac{1}{2}\eta^2\Delta \right)\epsilon(\eta,x) \label{gaugePsiA}
\ee
where $\epsilon$ is arbitrary.
Linear interactions with a background current can be added provided the current satisfies a ``continuity condition'': 
\be
S_{int}  =  \int d^4 x d^4\eta \delta'(\eta^2) J(\eta,x)\psi(\eta,x) \qquad \mbox{with} \qquad  \delta(\eta^2)\Delta J = 0. \label{eq:Continuity} 
\ee
Varying the action with respect to $\psi$, we obtain a new covariant equation of motion
\be
-\delta'(\eta^2) \Box_x \psi + \frac{1}{2} \Delta \left( \delta(\eta^2) \Delta \psi \right) = \delta'(\eta^2)J. \label{eomPsi1}
\ee
Exploring the basic physical content of the action \eqref{eq:BaseAction} and the equation of motion \eqref{eomPsi1} at both the classical and quantum level is the primary goal of this paper. 

Before describing further how physics is naturally encoded in the $\eta$-space, it is illuminating to transform to the space dual to $\eta$, where \eqref{eomPsi1} makes contact with familiar helicity equations of motion and currents. Expressing the equation of motion and continuity condition in terms of
$\psi(\omega,x)\equiv \int d^4\eta e^{i\eta\cdot\omega} \delta'(\eta^2)\psi(\eta,x)$ and 
$J(\omega,x)\equiv \int d^4\eta e^{i\eta\cdot\omega} \delta'(\eta^2)J(\eta,x)$, we obtain
\bea
\left( -\Box_x +(\omega\cdot \partial_x+i\kappa)\partial_{\omega}\cdot\partial_x - \frac{1}{2}(\omega\cdot\partial_x+i\kappa)^2\Box_{\omega} \right) \psi(\omega,x) &=& J(\omega,x), \label{eomPsi2} \\
\left( \partial_{\omega}\cdot\partial_x - \frac{1}{2}(\omega\cdot\partial_x +i\kappa)\partial_{\omega}^2 \right)J(\omega,x) & =& 0.
\eea
When $\kappa=0$, a restricted class of solutions to \eqref{eomPsi2} are the polynomial $\psi$ of the form 
\be
\psi = \phi(x) + \omega^{\mu}A_{\mu}(x)+\frac{1}{2}\omega^{\mu}\omega^{\nu}h_{\mu\nu}+\frac{1}{3!}\omega^{\mu}\omega^{\nu}\omega^{\rho}G_{\mu\nu\rho}... \label{eq:polyfields}
\ee
with currents 
$ J = J(x) + \omega^{\mu}J_{\mu}(x)+\frac{1}{2}\omega^{\mu}\omega^{\nu}\bar{J}_{\mu\nu}+...$.  With these restrictions (and $\kappa=0$), \eqref{eomPsi2} recovers the Fronsdal equations for high-spin particles \cite{Fronsdal:1978rb,Sorokin:2004ie}. Written in components, the equation of motion reduces to Klein-Gordon, Maxwell, and linearized Einstein equations for helicity 0,1, and 2 particles, and similarly for higher-helicity modes,
\bea
-\Box_x\phi -J &=& 0 \label{eq:KleinGordan} \nonumber \\
\omega^{\mu}\left(\Box_x A_{\mu} - \partial_{\mu}\partial\cdot A - J_{\mu} \right) &=& 0 \label{eq:Maxwell} \nonumber \\
\omega^{\mu}\omega^{\nu}\left(\Box_x h_{\mu\nu}-\partial_{\mu}\partial^{\sigma}h_{\nu\sigma}-\partial_{\nu}\partial^{\sigma}h_{\mu\sigma}+\partial_{\mu}\partial_{\nu}h^{\sigma}_{\sigma}-\bar{J}_{\mu\nu}\right) &=& 0 \qquad \dots, \label{eq:LinearGR} \nonumber
\eea
with the familiar conservation conditions $\partial^{\mu}J_{\mu}=0$, 
$\partial^{\mu}\left(\bar J_{\mu\nu}-\frac{1}{2} g_{\mu\nu}\bar J^{\sigma}_{\sigma}\right)=0$, etc.

We emphasize that generic $\eta$-space modes correspond to $\psi(\omega,x)$ that do \emph{not} admit the polynomial expansion \eqref{eq:polyfields} in $\omega$.  Indeed, there are a  continuous infinity of such modes (corresponding to suitably normalized functions on a ray) for every integer helicity. In fact, the polynomial solutions \eqref{eq:polyfields} are highly singular when transformed back to $\eta$-space and have ill-defined action \eqref{eq:BaseAction}.  Nonetheless, the transformation to $\omega$-space \emph{at the level of equations of motion} makes direct contact with the Fronsdal equations for helicity particles.   For $\kappa \ne 0$, the interpretation of polynomial solutions of \eqref{eomPsi2} is unclear.

\begin{figure}[!htbp]
\includegraphics[width=0.7\columnwidth]{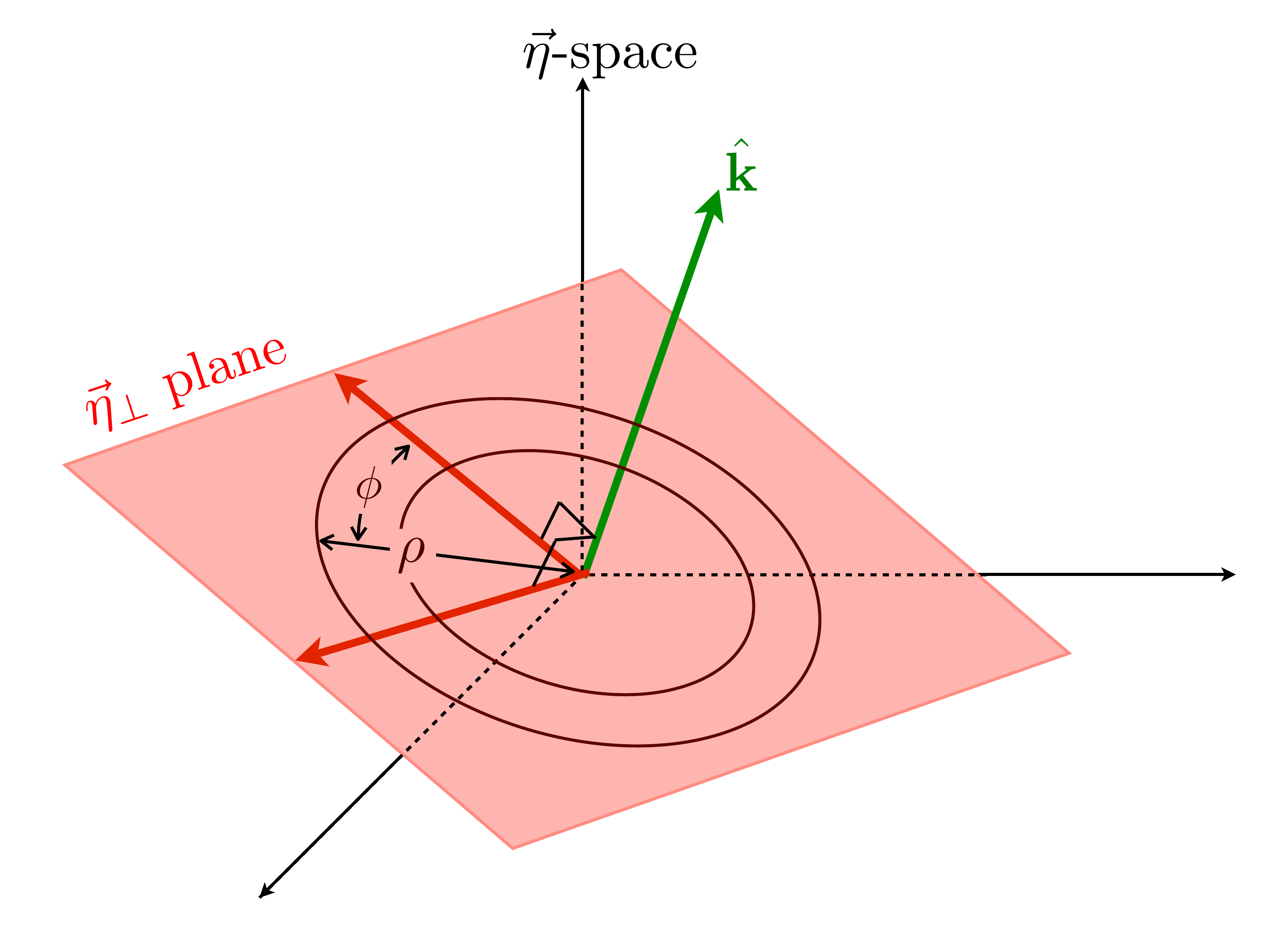}
\caption{Particle states in $D$-dimensional Minkowski space-time furnish representations of Wigner's Little Group ($E(D-2)$ for massless particles). 
The geometry of $\eta$-space is closely tied to that of the Little Group, as illustrated in the figure for massless particles in 3+1 dimensions.  Dynamical fields reside on the surface $\eta^2=0$, which can be parametrized by 3-vectors $\vec\eta$.  For every 3-momentum $\vec{\bf k}$, the field profile in the $\vec\eta.\vec{\bf k}$ direction can be fixed by a gauge choice.  Thus, physical information resides on the Euclidean $(D\!-\!2)$-plane transverse to $\vec{\bf k}$, shown shaded in the figure.  The massless Little Group $E(D\!-\!2)$ acts simply on this $\vec\eta_\perp$ plane. For non-zero $\kappa$, each concentric circle in $\vec\eta_\perp$ corresponds to a different CSP with a distinct spin-scale $\rho$. These can be interpreted as internal ($D$--2)-momenta of fixed magnitude.  
$\psi$'s localized in $\vec\eta_\perp$ correspond to eigenstates of Little Group translations (i.e. ``angle basis'' CSP  states \cite{Schuster:2013pxj}).  
When $\kappa=0$, Little Group translations act trivially and all circles on the $\vec\eta_\perp$ plane transform identically. The $\psi$ modes that rephase as $e^{in\phi}$ as they circle the origin are helicity-$n$ modes.\label{fig:transversePlane}}
\end{figure}

The $\eta$-space of \eqref{eq:BaseAction} has a very physical interpretation of its own: after fully gauge-fixing, it is associated with the Euclidean $(D\!-\!2)$-space on which the massless Little Group acts!  Although the action \eqref{eq:BaseAction} involves a field $\psi$ over the full $\eta$-space, it only depends on the value of $\psi$ in a first neighborhood of the $\eta$-null cone, or equivalently on two functions of a three-vector $\vec\eta$: $\psi$ on the $\eta$-null cone and its $\eta^0$-derivative.  Only one combination of these is dynamical.  The gauge invariance \eqref{gaugePsiA} further reduces the physical information: at momentum $k$, a choice of gauge fixes the derivative of $\psi$ in the $\vec\eta.\vec k$ direction (for $\kappa\neq 0$, this derivative is fixed but non-zero, echoing the non-transverse tensor description above).  This fully fixes the gauge, so that $\psi$ is specified by an arbitrary function on the $(D\!-\!2)$-plane orthogonal to $\vec k$, illustrated in Figure \ref{fig:transversePlane}.  

As we show in \S\ref{dof} for four dimensions, the $E(D\!-\!2)$ Little Group generators act naturally on this transverse $\vec\eta_\perp$ plane.   For non-zero $\kappa$, each concentric circle in $\vec\eta_\perp$ corresponds to a different CSP.  $\psi$'s localized in $\vec\eta_\perp$ correspond to eigenstates of Little Group translations (i.e. ``angle basis'' CSP  states \cite{Schuster:2013pxj}).  At $\kappa=0$, Little Group translations act trivially; the magnitude of $\vec\eta_\perp$ becomes irrelevant, and can be removed if desired by restricting \eqref{eq:BaseAction} to $\eta$-homogeneous $\psi$.  The helicity-$n$ mode of $\psi$ is a solution that rephases $n$ times as it circles the origin of the $\vec\eta_\perp$ plane.

The goal of this paper is limited --- to exhibit an action for CSPs, show that the free theory behaves
like a familiar healthy gauge theory, and quantize it by the ordinary canonical method.  Our primary aims are to 
show how CSP degrees of freedom emerge in the quantum theory, explain how to think about them in this new formalism, and sharpen open problems for future investigation.  We will include background currents that satisfy the continuity condition \eqref{eq:Continuity} to keep track of what the simplest type 
of linear interactions might look like and how to incorporate them, but even for $\kappa=0$ we have only a rudimentary understanding of interactions with matter in this formalism.  

Even this limited goal significantly improves on the state of the art.  No space-time action for CSPs has ever been proposed.  Wigner's equations of motion \cite{Wigner:1939cj,Bargmann:1946me,Bargmann:1948ck,Wigner:1963} and other Lorentz-covariant descriptions of CSPs are not amenable to quantization \cite{Iverson:1971hq,Abbott:1976bb,Hirata:1977ss}. These authors assumed strictly Little Group covariant fields, preventing their discovery of a gauge theory for CSPs.  Each paper encountered somewhat different obstructions to building properly causal fields and/or local Hamiltonians, underscoring that their negative results depend on the particular wavefunctions studied\footnote{Having classified all covariant wave equations for CSPs \emph{without} gauge redundancy \cite{Schuster:2013pxj}, it is now clear that none of these can be used to uniquely relate single-particle states to canonically quantized, local fields.}.

The key to our theory's viability is the introduction of a gauge redundancy in the description of CSPs.  The possibility that gauge invariance could play a crucial role in describing CSPs locally in space-time can be anticipated as they reduce in the $\rho\rightarrow 0$ limit to a tower of helicity-$h$ particles. This was already suggested in \cite{Bekaert:2005in}, where a gauge-invariant equation of motion very similar to \eqref{eomPsi2} was obtained from the high-spin limit of $(D+1)$-dimensional Fronsdal equations.  We have found a deformation of the action \eqref{eq:BaseAction} to space-like $\eta$ whose variation recovers that equation.  It is not clear whether this deformed action is as amenable to quantization as the form \eqref{eq:BaseAction}
\footnote{
The Fronsdal-like equation of motion of \cite{Bekaert:2005in} can be obtained by localizing \eqref{eq:BaseAction} on $\delta'(\eta^2+1)$ in our metric conventions and similarly modifying \eqref{gaugePsiA}.   This localization obstructs the restriction of \eqref{eq:BaseAction} to positive $\eta^0$ and the simple decomposition of $\psi$ in Section \ref{dynamical}, but alternate decompositions and subsequent quantizations may be possible.  Like our equations of motion, the Fronsdal-like equation of \cite{Bekaert:2005in} propagates CSPs of all $\rho$ unless further non-local restrictions are imposed.  An additional feature of \eqref{eomPsi1} and \eqref{eomPsi2} relative to those of \cite{Bekaert:2005in}, which may or may not be significant, is that \eqref{eomPsi1} and \eqref{eomPsi2} precisely recover the Fronsdal equations in the $\kappa \rightarrow 0$ limit.}. 

Even for $\kappa=0$, the simple action \eqref{eq:BaseAction} may be of interest for studying higher-spin fields \cite{Sorokin:2004ie,Bouatta:2004kk,Bekaert:2005vh,Fotopoulos:2008ka,Benincasa:2011pg}, though its eventual utility depends on how naturally matter interactions can respect the gauge invariance of \eqref{eq:BaseAction}.  
Until now, it was not known how to derive the Fronsdal equation 
from an action involving fields in a Minkowski spin-space (as originally formulated), and in such a simple manner.   
Known actions for the Fronsdal equations \cite{Fronsdal:1978rb} involve either an entourage of lower-rank auxiliary tensor fields for every helicity field or more sophisticated auxiliary spaces.  
In the geometrically intuitive $\eta$-space, the Fronsdal double traceless condition $(\partial_{\omega}^2)^2\psi=0$ on fields (and traceless condition on $\epsilon$) are replaced by the localization of the action, while the fields and gauge parameter are completely unconstrained. The shift to $\eta$-space requires that one embrace the full space of functions $\psi(\eta,x)$ or $\psi(\omega,x)$, not just the polynomial branch used by Fronsdal. 

Ultimately, we would like to know the physical origin of the action \eqref{eq:BaseAction}, whether other such actions exist, what kinds of matter sectors (if any) can furnish appropriately gauge-invariant interactions, and how \eqref{eq:BaseAction} generalizes to curved backgrounds and/or dynamical gravity.  Some of these issues will be discussed in \cite{ST:inprep}.  
The similarity of \eqref{eq:BaseAction} to known gauge theories hints at what structures may arise. 
The gauge transformation \eqref{gaugePsiA} is plausibly the linearization of a non-linear transformation. 
It would be interesting if this structure can be interpreted as the ``gauged'' version of a global symmetry,
perhaps of a matter sector that also lives in $\eta$-space. 
While our main motivation for developing this theory is to describe interacting CSPs, these questions apply equally to the description of helicity degrees of freedom with $\kappa=0$, and are quite familiar to those who have studied high-spin theories. Here, the basic ingredient that is widely thought neccesary for consistent high-spin interactions --- towers of helicity states --- occurs naturally 
and simply. 
Soft theorems for helicity particles \cite{Weinberg:1965rz,Weinberg:1964ew,Weinberg:1964ev,Porrati:2012rd,Benincasa:2007xk,Schuster:2008nh,Heinonen:2012km} and the helicity correspondence of CSPs \cite{Schuster:2} suggest that even in the presence of generic interactions, only the lowest-spin degrees of freedom couple strongly enough to mediate long-range forces.

The structure of this paper is as follows. In Section \ref{sec:action}, we elaborate on the action \eqref{eq:BaseAction}, introduce its homogeneous counterpart for $\kappa=0$, and present a simple path-integral quantization. Section \ref{dynamical} further explores the action, introducing a decomposition of $\psi$ into two functions of a three-vector $\vec\eta$ to clarify the dynamical field content.  Section \ref{canonical} constructs a Hamiltonian and canonically quantizes it in a Coulomb-like gauge.  Section \ref{dof} explores the action of the Little Group on fields of fixed momentum and shows definitively that the $\kappa\neq 0$ solutions to the equations of motion, and the particle states, are CSPs. Section \ref{singleCSP} shows how to define a $\eta$-space projective local Hamiltonian field theory for a single CSP of a prescribed $\rho$. We conclude and remark on the various open problems raised by this action in Section \ref{sec:conclusion}.  

%%%%%%%%%%%%%%%%%%%%%%%%%%%%%%%%%%%%
\section{Aspects of the Covariant Action}\label{sec:action}

In this paper we study and quantize the action 
\be
S_{free}  =  \frac{1}{2}\int d^4 x d^4\eta \left[ \delta'_+(\eta^2) (\partial_x \psi)^2 + \frac{1}{2}\delta_+(\eta^2)(\Delta\psi)^2 \right].\label{eq:BaseAction1}
\ee
Here we have written $\delta_+$ to explicitly localize to the \emph{positive} null cone $\eta^0=|{\vec\eta}|>0$; not doing so would simply double the number of degrees of freedom.  We will suppress the $\delta_+$ signs in the rest of the paper, but assume a localization to positive $\eta^0$ throughout this paper.  

This action has several features that suggest generalizations or variations.
The most obvious is to generalize both $x^\mu$ and $\eta^\mu$ to $D \neq 4$ spacetime dimensions.  In this case, the action will describe CSPs (and for $\kappa=0$ ordinary gauge fields) in higher dimensions.   
Though we work with $D=4$ for concreteness in this paper, all of our results generalize straightforwardly to higher dimensions.  Higher-dimensional continuous-spin representations are classified in \cite{Brink:2002zx}.   The localization of the action to the neighborhood of $\eta^2=0$ can also be modified by replacing both $\delta(\eta^2)$ and the $\eta^2$ in the gauge transformation \eqref{gaugePsiA} by $\eta^2\pm 1$.  The timelike case is very similar to \eqref{eq:BaseAction1}, while the spacelike one can no longer be restricted to positive $\eta^0$ while respecting Lorentz invariance.  We will not discuss either case further, and focus on the null-localized action. 
%Finally, mass terms that break the gauge symmetry can be added to \eqref{eq:BaseAction1}. This gives rise to a theory of massive integer spin particles with familiar looking massive equations of motion -- the extra propagating degrees of freedom are just the extra spin-polarizations that we'd expect. 

The action \eqref{eq:BaseAction1} does not describe a single CSP, but a half-line worth of CSPs
 with \emph{every} positive spin-scale $\rho$.  
For $\kappa=0$ this becomes a continuum of ``helicity towers''.  When $\kappa=0$, the action \eqref{eq:BaseAction1} is homogeneous under linear rescaling of $\eta$, suggesting a simple but likely non-unique modification that propagates only one state of every helicity.  We ``projectivize'' the action by restricting to $\psi$'s that are homogeneous functions of $\eta$ with weight $\frac{4-D}{2}$ in $D$ dimensions (weight $0$ in 3+1 dimensions), i.e. 
\be
\eta.\partial_\eta \psi(\eta,x) = \frac{4-D}{2} \psi(\eta,x),
\ee
so that the quantity in square brackets in \eqref{eq:BaseAction1} has projective weight $-D$.
The fields $\psi$ are functions of $\eta$-\emph{rays} (four-vectors $\eta^\mu$ and $\alpha \eta^\mu$ on the same ray are identified), which we can label by any representative $\eta$.  The natural measure over rays is
\be
d\mu(\eta) \equiv d^D\eta \delta(g(\eta)-1)
\ee
where $g(\eta)$ is {\it any} homogenous function of $\eta$ (not necessarily covariant) that is supported on every ray.  When this measure of weight $D$ is integrated against an integrand of weight $-D$, the action is independent of the choice of $g$.  
The final projectivized action is just
\be
S_{proj}=\frac{1}{2}\int d^4 x d\mu(\eta) \left[ \delta'(\eta^2) (\partial_x \psi)^2 + \frac{1}{2}\delta(\eta^2)(\partial_{\eta}\cdot\partial_x\psi)^2 \right].
\label{eq:projaction}
\ee

This action describes a single CSP with $\rho=0$, i.e. a single tower of integer helicities. The degrees of freedom before gauge-fixing are labeled by functions on the space of null rays, which has the topology of a $(D\!-\!2)$-sphere.  Gauge-fixing will reduce this to a function on a circle.  

The introduction of non-zero $\kappa$ breaks the homogeneity of the action \eqref{eq:BaseAction}, preventing us from using the same trick to obtain a theory of a single CSP.   
It is possible that CSPs can only be modeled in a simultaneously local and covariant manner if all $\rho$ are included simultaneously.  But the possibility remains open that a generalized homogeneity trick might suffice to isolate a single CSP.  In terms of creation and annihilation operators, the Hamiltonian we obtain from \eqref{eq:BaseAction} factorizes as an integral over $\rho$ and, as required by Lorentz invariance, different $\rho$ states do not mix under Lorentz transformations.  We have found a local ``gauge-fixed'' Hamiltonian (and generators for the full \Poincare algebra) for a single free CSP with fixed $\rho$, discussed in \S\ref{singleCSP}, but the gauge-fixing condition is spatially non-local and we do not yet know how to obtain it from a covariant action.  

\subsection{Simple Aspects of the Path-Integral}

Our main construction of the quantum theory in the rest of this paper will use canonical methods in a fully fixed gauge to keep unitarity manifest and clarify the propagating degrees of freedom.  But, as is familiar from gauge theories, this procedure obscures Lorentz covariance.  For this reason, we first present a derivation of covariant computational rules from the path integral.  The procedure is entirely standard, but we present a heuristic ``derivation'' in full (ignoring technical aspects such as the path integral measure and regularization) because the fields $\psi(\eta,x)$ are unfamiliar.  Our final result will be a familiar formula for computing correlation functions of fields in the presence of a background current $J(\eta,x)$. 

We use the familiar trick of introducing a background current $j(\eta,x)$ (not to be confused with $J$!) and computing the 
path integral partition function
\be
Z[j(\eta,x)] = \int D\psi e^{iS[\psi,j]}, \label{eq:PathIntegral}
\ee
where $j(\eta,x)$ is the background current coupled to $\psi$ as
\be
\int d^4 x d^4\eta \delta'(\eta^2) j(\eta,x)\psi(\eta,x).
\ee
It is useful to partially fix gauge to 
\be
\delta(\eta^2)\Delta\psi(\eta,x) = \delta(\eta^2)\omega(\eta,x),
\ee
where $\omega$ is an arbitrary function. 
This gauge can be reached by the transformation \eqref{gaugePsiA} with 
\be
\epsilon(\eta,x) = \frac{1}{\partial_x^2}(\omega-\Delta\psi).
\ee
To regulate the path integral \eqref{eq:PathIntegral}, we apply the usual Fadeev Popov technique to
 break up the path integral into a convergent piece that picks out a representative slice in field space 
 times a divergent (Lorentz invariant) factor that we throw away. 
We do this in the usual way by inserting the identity, 
\be
1= \int D\epsilon(\eta,x) \ \delta(G(\psi_{\epsilon})) \mbox{det}\left(\frac{\delta G(\psi_{\epsilon})}{\delta\epsilon}\right), \label{eq:PIidentity}
\ee
appropriatly into \eqref{eq:PathIntegral}, where
\be
G(\psi)=\delta(\eta^2)(\Delta\psi(\eta,x)-\omega(\eta,x)), \qquad 
\psi_{\epsilon}=\psi+(\eta\cdot\partial_x-\frac{1}{2}\eta^2\Delta)\epsilon.
\ee
The functional determinant factor $\mbox{det}\left( \delta(\eta^2)\partial_x^2 \right)$ 
is independent of $\epsilon$ and $\psi$, so we throw it away. 
We can perform a gauge transformation in the action to $\psi_{\epsilon}$ and 
shift the dummy variable of the path integral to $\psi_{\epsilon}$ to obtain 
\be
\left(\int D\epsilon\right) \int D\psi \ \delta(G(\psi)) \ e^{iS[\psi,j]},
\ee
and drop the divergent (Lorentz-invariant) factor $\int D\epsilon$ associated with integrating over
gauge-equivalent configurations. 
Following standard techniques, we use a normalized distribution in $\omega$ 
\be
\int D\omega \ \delta(G_{\omega}(\psi)) \ e^{-\frac{i}{2\xi}\int d^4x d^4\eta \delta(\eta^2)\omega^2}, 
\ee
parametrized by a free parameter $\xi$, instead of the factor \eqref{eq:PIidentity}.  Inserting this into the partition function and integrating over $\omega$ we obtain 
\be
Z[j(\eta,x)] = \int D\psi e^{iS_{\xi}[\psi,j]}, \label{eq:xiPathIntegral}
\ee
where
\bea
S_{\xi}[\psi,j,] & = & \int d^4 x d^4\eta \left( \frac{1}{2} \delta'(\eta^2) (\partial^\mu_x \psi)^2 + \frac{1}{4}(1-\xi^{-1}) \delta(\eta^2) \left( \partial_\eta.\partial_x\psi\right)^2 + \delta'(\eta^2)j\psi \right) \label{eq:xiAction} \\
& = & \int d^4 x d^4\eta\delta'(\eta^2) \left(
\frac{1}{2} \psi \hat{D}^{-1}_{\xi}(\eta,x)\psi
+ j\psi \right). 
\eea
where the kernel $\hat{D}^{-1}_{\xi}$ is
\be
\hat{D}^{-1}_{\xi}(\eta,x) = \left(-\partial_x^2 +(1-\xi^{-1})\eta\cdot\partial_x \partial_{\eta}\cdot\partial_x - \frac{1}{2}(1-\xi^{-1})\eta^2(\partial_{\eta}\cdot\partial_x)^2 \right).
\ee 
To complete the square in the action, we must find a Green's function $\hat{D}_{\xi}(\eta,\eta';x-y)$ that inverts $\hat{D}^{-1}_\xi$ in the 1-st neighbourhood of the $\eta$-space delta functions, i.e. 
\be
\delta^{\prime}(\eta^2)\delta^{\prime}(\eta'^2)\hat{D}^{-1}_{\xi}(\eta,x)\hat{D}_{\xi}(\eta,\eta';x-y) = i\delta^{\prime}(\eta^2)\delta^4(x-y)\delta^4(\eta-\eta'),
\ee
and shift $\psi$ to
\be
\psi(\eta,x) \rightarrow \psi'(\eta,x)=\psi(\eta,x)-i\int d^4y d^4\eta' \delta'(\eta'^2) \hat{D}_{\xi}(\eta,\eta',x-y)j(\eta',y).
\ee
The $\xi=1$ Green's function is particularly simple: 
\be
\hat{D}_{\xi=1}(\eta,\eta';x-y) = \delta^4(\eta-\eta')D_F(x-y),
\ee
where $D_F(x-y)$ is the usual Feynman propogator.
In terms of the shifted $\psi$ we obtain
\bea
&& \int d^4 x d^4\eta \delta'(\eta^2) \left(
\frac{1}{2} \psi \hat{D}^{-1}_{\xi}(\eta,x)\psi \right) \nonumber \\
+&&
\int d^4 x d^4\eta d^4y d^4\eta' \delta'(\eta^2)\delta'(\eta'^2) \left( j(\eta,x)\hat{D}_{\xi}(\eta,\eta';x-y)j(\eta',y) \right). 
\eea
so that the generating function takes the familiar Gaussian form 
\be
Z_{\xi}[j] = e^{i\int d^4 x d^4\eta d^4 y d^4\eta' \delta'(\eta^2)\delta'(\eta'^2) \left(j(\eta,x)\hat{D}_{\xi}(\eta,\eta';x-y)j(\eta',y) \right)}
\ee
where we have dropped a $j$-independent Gaussian factor.

Correlation functions in $\xi$-gauge can now be simply computed.  
For example, the propagator is just
\be
\delta^{\prime}(\eta^2)\delta^{\prime}(\eta'^2)\bra{\mbox{vac}} T \psi(\eta,x)\psi(\eta',y)\ket{\mbox{vac}} = \frac{\delta}{\delta j(\eta,x)}\frac{\delta}{\delta j(\eta',y)} Z_{\xi}[j]_{j=0} = \delta^{\prime}(\eta^2)\hat{D}_{\xi}(\eta,\eta';x-y),
\ee
as expected.  If we couple $\psi$ to a current $J(\eta,x)$ built out of matter fields, the matter correlation functions, and contributions from $\psi$ scattering off $J$ can be computed similarly.  The main point of exhibiting the above was to show how covariant computational rules, despite the unfamiliar 
$\eta$-space, are straightforward to derive. 
The key challenge at this point is building a $J(\eta,x)$ from matter fields that non-trivially satisfy the continuity condition (beyond the trivial ``scalar'' currents $J(\eta,x)=J(x)$ when $\kappa=0$).  As we discuss in \cite{ST:inprep} and highlight in \S\ref{ssec:examples}, it is easy enough to insert currents in the Hamiltonian that yield covariant results, but less clear how to include single tensor currents (i.e. $J^{\mu}(x)$) built out of single tensor matter fields (i.e. $\phi(x)$) in the form of a smooth and covariant $J(\eta,x)$.
%This suggests that perhaps \eqref{eq:BaseAction} must be generalized to allow more general couplings to matter.  

%%%%%%%%%%%%%%%%%%%%%%%%%%%%%%%%%%%%%%%%%%%%%%
\section{Dynamical and Non-Dynamical Fields}\label{dynamical}

In this section we characterize the dynamical degrees of freedom of the action \eqref{eq:BaseAction}.  
Constructing a Hamiltonian (in the next Section) will force us to single out a time direction, thereby breaking manifest Lorentz covariance. 
Doing so from the beginning of the analysis will clarify which degrees of freedom are dynamical and which are not (as discussed in this section), 
and will later allow us to sharpen how the CSP particle degrees of freedom emerge.  
Though the fields are organized quite differently from the usual tensor gauge fields, when we specialize to the $\kappa=0$ action 
and perform tensor projections, we will recover a theory of propagating helicity degrees of freedom. 
For the projectivized action, one mode of each integer helicity will propagate.

\subsection{Decomposing $\psi$ in a neighborhood of the null cone}

Because the action only depends on the field $\psi(\eta,x)$ in a first-order neighborhood of the positive null $\eta$-cone, it is useful to expand in a decomposition 
\be
\psi(\eta,x) = \psi_0(\vec\eta,x) + (\eta^0-r)\psi_1(\vec\eta,x)+(\eta^0-r)^2\psi_2(\vec\eta,x)+...\label{psiI_decomp}
\ee
where $r=\sqrt{\vec\eta^2}$ and $\partial_{\eta^0} \psi_i = 0$.  The terms $\psi_2$ and higher in the expansion do not contribute to the action at all. 
It will prove useful to group the first two terms as \be
\psi(\eta,x) = A(\vec\eta,x) + \frac{\eta^0}{r} B(\vec\eta,x)+{\cal O}(\eta^0-r)^2,\label{AB_decomp}
\ee
where $A=\psi_0 - r \psi_1$ and $B=r \psi_1$.  
In the homogeneous case with $\eta.\partial_\eta \psi = 0$, we also have $\vec\eta.\grad_\eta A = \vec\eta.\grad_\eta B = 0$.  
Expanding the current similarly as
\be
J(\eta,x) = J_A(\vec\eta,x) + \frac{\eta^0}{r} J_B(\vec\eta,x)+{\cal O}(\eta^0-r)^2,\label{JAB_decomp} 
\ee
and using the identity 
\be
\delta^{\prime}_+(\eta^2) = \delta_+(\eta^2)\frac{1}{2(\eta^0)^2}\left(1-\eta^0\partial_{\eta^0}  \right),
\ee
we can rewrite the action as
\be
S   =  \int d^4 x d^4\eta \frac{1}{2r^2} \delta_+(\eta^2)\left[\tfrac{1}{2}\left((\partial_x A)^2 - (\partial_x B)^2\right) + \tfrac{1}{2} r^2 (\tfrac{1}{r}\dot B +r \deltabar \tfrac{1}{r}B + \deltabar A)^2 + (J_A A - J_B B)\right].
\ee
where $\deltabar \equiv -\grad_\eta.\grad_x+\kappa$ (this is just the ``spatial'' part of $\Delta$, i.e. $\Delta = \partial_x^0 \partial_\eta^0 + \deltabar$).  
The two terms quadratic in $\dot B$ (from the second and third terms respectively) cancel one another, so that $B$ is non-dynamical while $A$ has a canonical kinetic term.  To further simplify this expression, we localize the $\eta^0$ integral on $\delta_+(\eta^2)$, then use integration by parts to simplify the quadratic terms in $B$ to a single term: 
\be
S  = \int d^4 x \frac{d^3\eta}{4 r^3} \left[\tfrac{1}{2}\left((\partial_x A)^2 + (r^3 \deltabar\tfrac{1}{r^2} B)^2 + r^2 (\deltabar A)^2\right) + r \dot B \deltabar A + r^3 (\deltabar\tfrac{1}{r}B)\,\deltabar A + J_A A - J_B B\right].\label{ABaction}
\ee
The $\eta$-integration yields a bounded Lagrangian density provided that integrals of $A$, $B$, $A J_A$ and $B J_B$ over the solid angle of a radius-$r$ sphere vanish faster than $1/r$ both as $r\rightarrow 0$ and as $r\rightarrow \infty$.
%\scratch{SHOULDN'T THIS BE VANISHING FASTER THAN A CONSTANT??} 
Variation of the action \eqref{ABaction} yields the two Lagrangian equations of motion 
\bea
(\Box - r^3 \deltabar \tfrac{1}{r} \deltabar) A & = & J_A + r^3 \deltabar \tfrac{1}{r^2} \dot B + r^3 \deltabar^2 \tfrac{1}{r} B, \label{Aeom}\\
r \deltabar r^3 \deltabar \tfrac{1}{r^2} B & = & J_B + r \deltabar \dot A - r^2 \deltabar^2 A \equiv {\cal J}_B, \label{Beom}
\eea
again making clear that $B$ is non-dynamical.  These are the same as the covariant equations of motion \eqref{eomPsi1} after 
expanding using \eqref{AB_decomp} and integrating over $\eta^0$ to localize to the null $\eta$-cone. 

It is not difficult to invert \eqref{Beom} and solve for $B$ by constructing a suitable Green's function $G_D(\vec\eta,\vec\eta',x-x')$ that satisfies
\be
D G_D(\vec\eta,\vec\eta',x-x') = \bar\delta(\vec\eta,\vec\eta',x-x') = 4 r^3 \delta^{(3)}(\vec\eta-\vec\eta')\delta^{(3)}(x-x')
\ee
where $D \equiv r \deltabar r^3 \deltabar \tfrac{1}{r^2}$. 
The solution for $B$ is then
\be
B(\vec\eta,\vec x) = \int \frac{d^3\vec\eta'}{r'^3}d^3{\vec x} \,  G_D(\vec\eta,\vec\eta',\vec x- \vec x') {\cal J}_B(\vec\eta',x').\label{Bsolution}
\ee
An explicit formula for $G_D$ is derived in Appendix \ref{app:greens}.
  
%%%%%%%%%
\subsection{Gauge Fixing and Gauge-Invariants}

Although the $A$ field is dynamical, it still possesses a gauge redundancy, so that only a subset of the information in $A(\vec\eta,x)$ is physical.  Using \eqref{psiI_decomp}, and decomposing the gauge parameter as $\epsilon(\eta,x) = \epsilon_0(\vec\eta,x) + {\cal O}(\eta^2)$, the gauge transformation \eqref{gaugePsiA} acts on $A$ and $B$ as
\be
\delta A = (\vec\eta.\grad_x + r^2 \deltabar) \epsilon_0  = r^3 \deltabar \tfrac{1}{r} \epsilon_0 \qquad \delta B = r\dot \epsilon_0 - r^2\deltabar \epsilon_0\label{ABgauge}
\ee
and is determined by the single function $\epsilon_0(\vec\eta,x)$ on the null $\eta$ cone.  These variations satisfy our earlier boundedness requirements provided that $r \epsilon_0$ has a bounded integral over the large-$\eta$ boundary.

Because $\deltabar$ is the only $\eta$-derivative that appears in our equations of motion and gauge variations, they are most conveniently studied by Fourier transforming to 3-momentum space ($\grad_x = i \vec {\bf k}$, $\deltabar = -i \vec{\bf  k}.\grad_\eta+\mu$) and introducing $\vec{\bf k}$-centric cylindrical coordinates $(z,\eta_\perp,\phi)$ for $\vec\eta$ where $z = \vec\eta\cdot\hat{\bf k}$, $\eta_\perp = |\vec\eta\cross\hat{\bf k}|$, and $\phi$ is an angle in the plane transverse to ${\bf k}$.  In these coordinates, $\deltabar = -i |{\bf k}| \partial_z+\kappa$.

Note that the variation of $e^{i\kappa \vec\eta.{\bf k}/|{\bf k}^2|} A/r^3$ in \eqref{ABgauge} is a total $\eta$-derivative of a bounded function $e^{i\kappa \vec\eta.{\bf k}/|{\bf k}^2|} \epsilon_0/r $.  This implies that its line integral in the $\vec\eta.{\bf k}$ direction is gauge-invariant.  In our ${\bf k}$-centric cylindrical coordinates this is simply 
\bea
f(\eta_\perp,\phi,\vec{\bf k}, t) &=& \frac{1}{2} \int_{-\infty}^{+\infty} \frac{dz}{r^3} A(z,\eta_\perp,\phi,\vec{\bf k}, t) e^{i\kappa z/|{\bf k}|}. \label{fGI}
%& = & \frac{d^3\eta'}{r'^3} \delta_\perp^{(2)}(\vec\eta-\vec\eta') A(\vec\eta',\vec{\bf k}, t) e^{i\mu \vec\eta'.{\bf k}/|{\bf k}^2|} 
\eea
The function $f$ is neither Lorentz-covariant nor local in position space, and so it is not an especially natural way of parametrizing degrees of freedom, but does enable us to count them.  For every momentum ${\bf k}$, $f$ has a function on a plane's worth of information (the $E(2)$ plane of Figure \ref{fig:transversePlane}). 

In the $\kappa=0$ projective action, the restriction to homogeneous $A$ implies that we need only specify $f$ on one circle, say $\eta_\perp=1$, to determine $f(\eta_\perp,\phi)$ for all $\eta_\perp$.  Thus, in this case $f$ contains only a function on a unit circle's worth of information.  This can be decomposed into Fourier coefficients for all integers $n$. Under rotations about $\hat k$ that map $\phi \rightarrow \phi+\theta$, the Fourier coefficients rephase by $e^{in\theta}$ so that they should be associated with helicity-$n$ degrees of freedom.  
  
It is possible to gauge-transform a generic $A(\vec\eta,x)$ into an $A'$ with $\deltabar A'=0$, and doing so fully fixes the gauge.  
This is easily seen in $\vec{\bf k}$-centric cylindrical coordinates.  Configurations $A_1$ and $A_2$ are gauge-equivalent if 
\bea
e^{i\kappa z/|{\bf k}|} \epsilon(z,\eta_\perp,\phi) &= &r_z \int_{-\infty}^z \frac{dz'}{r'^3} (A_2(z')-A_1(z')) e^{i\kappa z'/|{\bf k}|} \\
&= &  - r_z \int_z^{+\infty} \frac{dz'}{r'^3} (A_2(z')-A_1(z')) e^{i\kappa z'/|{\bf k}|} + 2 r_z (f_2 - f_1) \label{epsilonA12}
\eea
(where $f_1$ and $f_2$ are given by \eqref{fGI} with gauge fields $A_1$ and $A_2$) is sufficiently bounded as $z\rightarrow \pm \infty$.  Using the bounded  growth of $A_{1,2}$ as $z\rightarrow \pm \infty$, the condition  $f_1(\eta_\perp,\phi) = f_2(\eta_\perp,\phi)$ is both necessary and sufficient to guarantee that $\epsilon(z,\phi)$ from \eqref{epsilonA12} is a valid gauge transformation.  
Thus, field configurations $A$ with the same $f$ are gauge-equivalent to one another, while solutions with different $f$ are inequivalent.   The $\deltabar A = 0$ gauge corresponds to a flat profile $A(z,\eta_\perp,\phi) = f(\eta_\perp,\phi)$ when $\kappa=0$, and to a fixed profile $A(z,\eta_\perp,\phi) = e^{i\kappa z/|{\bf k}|} f(\eta_\perp,\phi) $ for all $z$ when $\kappa\neq 0$.  We note also that any gauge transformation $\epsilon$ that falls as $z\rightarrow \pm \infty$ will necessarily change $A$, so that $\deltabar A=0$ fully fixes gauge.  

In $\deltabar A = 0 $ gauge, the Lagrangian equations of motion simplify considerably.  In particular, $B$ is not sourced by $A$ so the effective source in \eqref{Bsolution} is just ${\cal J}_B = J_B$.  Since $A(z,\eta_\perp,\phi) e^{-i\mu z/|{\bf k}|} = f(\eta_\perp,\phi)$ in this gauge, we can take the appropriately weighted integral of \eqref{Aeom} to obtain the simple $\deltabar A  = 0 $ equation of motion 
\bea
\Box A(z,\phi)  = \f{1}{2} \int_{-\infty}^{+\infty} \frac{dz'}{r'^3} J_A(z',\phi) e^{i\kappa (z'-z)/|{\bf k}|}.
\eea
Note that the right hand side satisfies the $\deltabar$-condition. 

We note that the gauge condition $\deltabar A = (- \grad_x . \grad_\eta +\kappa ) A$ is quite reminiscent of the Coulomb gauge in QED, and indeed the remainder of our discussion will continue to parallel the textbook canonical quantization of Coulomb-gauge QED as presented in e.g. \cite{Bjorken:1965zz, WeinbergQFT}.  We can heuristically think of $\grad_{\vec\eta}^{i_1} \dots \grad_{\vec\eta}^{i_n} A(\vec\eta,x)$ as encoding the same information as a rank-$n$ symmetric tensor $V^{i_1 \dots i_n}$ subject to the Coulomb-gauge condition $\grad_{x}^{i_1} V^{i_1 \dots i_n} = 0$.  The time-components of a rank-$n$ covariant vector $A^0$, $h^{0i}$, etc. are non-dynamical and are encoded in successive $\vec\eta$ gradients of the field $B(\vec\eta,x)$.  From helicity 2 onward, the usual tensor fields also have terms with at least \emph{two} timelike indices, but in traceless gauge they carry no information that could not be obtained from the previously considered components.  Thus we suspect that the packaging of degrees of freedom in our $\eta$-space action is most similar to that of traceless gauge fields, even though the covariant equation of motion is Fourier-conjugate to the \emph{double-traceless} Fronsdal equation. This suggests the possibility that a generalization of the action \eqref{eq:BaseAction}, with even larger gauge symmetry, may exist. 

%%%%%%%%%%%%%%%%%%%%%%
\section{Canonical Quantization}\label{canonical}

Though we can derive a Hamiltonian from the action \eqref{ABaction}, it is slightly simpler to first integrate by parts so that the action is independent of $\dot B$, then regroup terms into the form 
\be
L  = \int d^3 x \frac{d^3\vec\eta}{4 r^3} \left[\frac{1}{2}\left(\left(\dot A - r^3 \deltabar\tfrac{1}{r^2} B\right)^2  -(\grad_x A)^2  + r^2 (\deltabar A)^2\right) + r^3 (\deltabar\tfrac{1}{r}B)\,\deltabar A + J_A A - J_B B\right],\label{ABaction2}
\ee
To facilitate later contact with the homogeneous $\kappa=0$ action, we will use the homogeneous measure $d^3\vec\eta/(4r^3)$ throughout this section, as well as the $\vec\eta$-homogeneity-preserving functional derivative
\be
\frac{\delta A(\vec\eta,\vec x)}{\delta A(\vec\eta',\vec x')} \equiv \bar\delta(\vec\eta,\vec\eta') \equiv 4 |{\vec\eta}|^3 \delta^{(3)}(\vec\eta - \vec\eta') \delta^3(\vec x - \vec x').
\ee
Conventions for the homogeneous action are given in Appendix \ref{app:homog} and yield results identical to those in this section except for the measure substitution $\frac{d^3\vec\eta}{4r^3}\rightarrow D^2\vec\eta$.
In constructing the Hamiltonian, the Poisson bracket is defined in the usual way as
\be
\{ O_1(\vec\eta,\vec x) , O_2(\vec\eta',\vec x') \} \equiv \int d^3 x''\frac{d^3\vec\eta''}{4 r''^3} \left[ \frac{\delta O_1}{\delta A''}\frac{\delta O_2}{\delta \Pi_{A}''}
-\frac{\delta O_1}{\delta \Pi_{A}''}\frac{\delta O_2'}{\delta A} \right] + (B-terms)
\ee
where the shorthand $O'$ denotes evaluation at $(\vec\eta',\vec x')$, 
so that
\be
\{ A(\vec\eta,\vec x) , \Pi_{A}(\vec\eta',\vec x') \} = \bar\delta(\vec\eta,\vec\eta',x-x').
\ee

We follow Dirac's procedure to find a family of ``extended'' Hamiltonians for \eqref{ABaction2} \emph{before} fixing gauge \cite{DiracLectures}.  The canonically conjugate variables are 
\be
\Pi_{A} = \frac{\delta L}{\delta \dot A} = \dot A - r^3 \deltabar\tfrac{1}{r^2} B \qquad \Pi_B = 0
\ee
(both are reminiscent of the QED relations $\vec\Pi = \dot {\vec A} - \grad A^0$, $\Pi^0 = 0$).  The latter $\Pi_B(\vec\eta,\vec x) = 0$ encodes a field's worth of primary constraints (i.e. one constraint for each $\vec\eta$ and $\vec x$).  The canonical Hamiltonian is therefore 
\bea
H_c & = &\int d^3 x \frac{d^3\vec\eta}{4 r^3} \Pi_A(\vec\eta,\vec x) A(\vec\eta,\vec x) - L  \label{Hcan} \\
& = & \int d^3 x \frac{d^3\vec\eta}{4 r^3} \left[ \tfrac{1}{2}\Pi_A^2 + \tfrac{1}{2}(\grad A)^2 - \tfrac{1}{2}(r \deltabar A)^2+ r^3 (\Pi_A \deltabar(\tfrac{1}{r^2} B) - (\deltabar\tfrac{1}{r}B)\,\deltabar A) - J_A A + J_B B \right]. \nonumber
\eea
The first two terms are just familiar kinetic terms for the dynamical field $A$.
The primary constraint $\phi_1: \Pi_B = 0$ has non-vanishing Poisson bracket with the Hamiltonian, implying an additional secondary constraint
\be
\phi_2 = \{\phi_1, H\} = - \frac{\delta H}{\delta B} = - (J_B + r\deltabar \Pi_A - r^2 \deltabar^2 A).
\ee
This constraint \emph{does} have vanishing Poisson bracket with the canonical Hamiltonian (more precisely, the Poisson bracket is proportional to the current continuity condition, which we assume to vanish), so the constraint system closes.  It is obvious that $\{\phi_1(\vec\eta,\vec x), \phi_i(\vec\eta',\vec x')\} = 0$ for $i=1$ or 2.  The last Poisson bracket of constraints,
\bea
\{\phi_2(\vec\eta,\vec x), \phi_2(\vec\eta',\vec x')\} & = & \{J_B(\vec\eta,\vec x), J_B(\vec\eta',\vec x')\} + (r' \deltabar_{\eta'}  - r \deltabar_{\eta}) \deltabar_{\eta'}\deltabar_{\eta} \bar\delta(\vec\eta,\vec\eta',\vec x-\vec x') \\
& =& \{J_B(\vec\eta,\vec x), J_B(\vec\eta',\vec x')\}
\eea
also vanishes so long as $J_B$ has vanishing equal-time Poisson bracket with itself, so that the constraint system is first-class.  First class constraints are generally associated with generators of gauge transformations.  To demonstrate this connection here, we first introduce a linear combination of $\phi_1$ and $\phi_2$ weighted by two arbitrary functions of $\vec\eta$ and $\vec x$:
\be
\phi[f,g] = \int d^3 x \frac{d^3\vec\eta}{4 r^3} \left[ f(\vec\eta,\vec x) \phi_1(\vec\eta,\vec x) + g(\vec\eta,\vec x) \phi_2(\vec\eta,\vec x) \right].  
\ee
We find
\be
\{A, \phi[f,g]\} = r^3 \deltabar \tfrac{1}{r^2} g \quad
\{\Pi_A, \phi[f,g]\} = r^3 \deltabar^2 \tfrac{1}{r} g \quad
\{B, \phi[f,g]\} = f.
\ee
Using $\Pi_A = \dot A - r^3 \deltabar \tfrac{1}{r^2} B$, we find $f = \dot g - r^2 \deltabar \tfrac{1}{r} g$, so that $g/r$ plays the role of the gauge parameter $\epsilon_0$.  
Thus the first-class constraints $\phi_1$ and $\phi_2$ precisely generate the gauge transformations \eqref{ABgauge}.

Following Dirac, we can consider a family of extended Hamiltonians parametrized by functions $\alpha_1$ and $\alpha_2$ of the canonical variables, 
\bea
H_e & = & H_c + \int d^3 x \frac{d^3\vec\eta}{4 r^3} \left[ \alpha_1(\vec\eta,\vec x) \phi_1(\vec\eta,\vec x) + \alpha_2(\vec\eta,\vec x) \phi_2(\vec\eta,\vec x) \right]  \\
& = & \int d^3 x \frac{d^3\vec\eta}{4 r^3} \, \bigg[ \tfrac{1}{2}\! \left( \Pi_A^2 + (\grad A)^2 - (r \deltabar A)^2\right) - J_A A + \alpha_1 \Pi_B \nonumber \\
& & \qquad \qquad  \qquad \qquad \qquad \qquad+ (B - \alpha_2) (J_B + r\deltabar \Pi_A - r^2 \deltabar^2 A) \bigg]. \label{Hext}
\eea
Each choice of $\alpha_{1,2}$ generates a distinct but gauge-equivalent time evolution.  By choosing $\alpha_1 = 0$ and $\alpha_2 = B - \tilde \alpha_2$, where $\tilde \alpha_2$ is independent of $B$ and $\Pi_B$, we obtain a Hamiltonian exclusively for the dynamical fields $A$ and $\Pi_A$.  Of particular interest is the Hamiltonian that preserves the gauge choice $\deltabar A = 0$.  To find such a Hamiltonian, one must solve $\{\deltabar A, H_{e}\} = 0$ for the parameters $\alpha_{1,2}$.  
\be
\{\deltabar A, H_{e}\} = \tfrac{1}{r}\left( r \deltabar \Pi_A + r \deltabar r^3 \deltabar \tfrac{1}{r^2} \tilde \alpha_2 \right) \simeq \tfrac{1}{r} \left( D\tilde \alpha_2 - J_B\right),
\ee
where $\simeq$ denotes ``weak equality'' up to the constraints $\phi_{1,2}$ and the gauge condition $\deltabar A$.  We recognize 
$D \equiv r \deltabar r^3 \deltabar \tfrac{1}{r^2}$ as the differential operator inverted by the Green's
function $G_D$ in \eqref{eq:GDequation}.
Thus the choice $\tilde\alpha_2 = G_D J_B$ is compatible with the Coulomb-like gauge condition $\deltabar A = 0$.  
With this substitution, we obtain the unique Hamiltonian that generates time-evolution compatible with $\deltabar$=0 gauge:
\be
H = \int d^3 x \frac{d^3\vec\eta}{4 r^3} \left[ \tfrac{1}{2} \left( \Pi_A^2 + (\grad A)^2 - (r \deltabar A)^2\right) - J_A A + (G_D J_B) (J_B + r\deltabar \Pi_A - r^2 \deltabar^2 A) \right].\label{Hcoulomb}
\ee
The two constraints in this gauge are $\phi_2 = - (J_B + r\deltabar \Pi_A - r^2 \deltabar^2 A) = 0$ and $\chi = r \deltabar A  = 0$.  
These have non-vanishing Poisson brackets with each other, and are therefore \emph{second-class} constraints with Poisson bracket
\be
\{\phi_2(\vec\eta,\vec x), \chi(\vec\eta',\vec x')\} = D \bar\delta(\vec\eta,\vec\eta',\vec x - \vec x'). \label{2ndclass}
\ee

It is now completely straightforward to canonically quantize the classical theory defined by the master action \eqref{eq:BaseAction} 
using Dirac brackets. We provide the computation of the appropriate Dirac brackets of the canonical fields $A$ and $\Pi_A$
in Appendix \ref{app:dirac}. 
The resulting {\it canonical commutation relations} for the fields $A$ and $\Pi_A$ are obtained by multiplying by $i$:
\begin{align}
\left[A(\vec\eta,x), A(\vec\eta',x')\right] & =  \left[\Pi_A(\vec\eta,x),\Pi_A(\vec\eta',x')\right] = 0 \\
\left[A({\vec\eta},x),\Pi_A(\vec\eta',x')\right] & = 
i\bar\delta_{\deltabar}(  \vec\eta,\vec\eta', \vec x - \vec x' )\label{DiracBrackets}
\end{align}
where 
\be
\bar\delta_{\deltabar}(\vec\eta,\vec\eta',x-x') \equiv \int d^3k e^{i{\bf k}.{\bf x}} 2 \eta_\perp^2 \delta^{(2)}_\perp(\vec\eta-\vec\eta') e^{i \kappa (\vec\eta'-\vec\eta).{\bf k}/|{\bf k}|^2},\label{deltaperpDef}
\ee
$\eta_\perp^2 = \frac{|\vec\eta\cross {\bf k}|^2}{|{\bf k}|^2}$, and $\delta^{(2)}_\perp$ is the two-dimensional delta-function in the $\vec\eta_\perp$ plane transverse to ${\bf k}$.
As usual, this behaves as a ``transverse'' delta-function that projects onto the space consistent with the constraints. 
In particular, $\deltabar \delta_{\deltabar} = 0$ and for any $X$ with  $\deltabar X(\vec\eta,x) = 0$,
\be
\int \frac{d^3\eta'}{4 r'^3} d^3 {\bf x'} \bar\delta_{\deltabar}(  \vec\eta,\vec\eta', {\bf x} - {\bf x'}) X(\vec\eta',x')
=  X(\vec\eta,x),
\ee
which can be obtained simply by performing the integral over $\eta.{\hat {\bf k}}$ in \eqref{deltaperpDef} and using the gauge condition.   

So far, the quantization of the classical theory given by our action \eqref{eq:BaseAction} has worked essentially 
like quantizing QED, and as we'll see, the $\deltabar$=0 gauge is just like Coulomb-gauge.
In the presence of a background current, one further step is usually performed, namely a change of variables from the canonical momentum $\vec\Pi$ (which satisfies a non-homogeneous secondary constraint $\grad.\vec\Pi = J^0$) to a ``transverse momentum'' $\vec\Pi_\perp$ satisfying $\grad.\vec\Pi_\perp=0$.  This change of variables has the important consequence that $\vec\Pi_\perp$, when canonically quantized using its Dirac bracket, commutes with $J^0$ and hence with matter fields, while $\vec\Pi$ does \emph{not} commute with $J^0$ and hence must not commute with all matter fields.  
A highly analogous situation arises here: The commutators between any matter fields in $J_B$ and $\Pi_A$ receive non-zero contributions.  
But if we define
\be
\Pi_\perp \equiv \Pi_A + r^3 \deltabar \tfrac{1}{r^2} D^{-1} J_B 
\ee
then $\phi_2$ can be expressed as $r \deltabar \Pi_\perp -r^2 \deltabar^2 A = 0$, which has vanishing commutator with $J_B$ and 
hence permits trivial commutation relations between $\Pi_\perp$ and any matter fields. 
The commutation relations of $\Pi_\perp$ with itself and with $A$ are unchanged from \eqref{DiracBrackets}. 

In terms of $\Pi_\perp$, 
\be
H = H_{free}+V_{int} \label{Hfinal}
\ee
where
\be
H_{free} = \int d^3 x \frac{d^3\vec\eta}{4 r^3} \left[ \tfrac{1}{2} \Pi_\perp^2 + \tfrac{1}{2} (\grad A)^2 \right] \label{Hfree}
\ee
and 
\be
V_{int} = \int d^3 x \frac{d^3\vec\eta}{4 r^3} \left[ - J_A A + \tfrac{1}{2} J_B G_D J_B \right] \label{Vint}
\ee
This is a very familiar looking free Hamiltonian with a coupling to a background current! 

At this point, it's straightforward to pass to the interaction picture and use standard perturbative techniques to calculate S-matrix elements for CSP scattering in the presence of the background current $J$. Likewise, we can work out the propagator in the usual fashion, and show that it's equivalent 
to the covariant $\xi$-propagator derived using path-integral methods earlier, provided the current satisfies the continuity condition. 

\subsection{A Few Examples}\label{ssec:examples}

The interaction energy associated with the background current $J_B$ --- the static part --- is just 
\bea
V_{static} &=& \int d^3 x \frac{d^3\vec\eta}{4 r^3} \tfrac{1}{2} J_B G_D J_B \\
&=& \int d^3{\bf k} \frac{d^3\vec\eta}{4r^3} \frac{d^3\vec\eta'}{4r'^3} 
|\eta_\perp|^2 \delta_\perp^{(2)}(\vec\eta-\vec\eta') e^{i\kappa (\vec\eta'-\vec\eta).{\bf k}/|{\bf k}|^2} \frac{(r \tilde{J}_B(\vec\eta,{\bf k})) (r' \tilde{J}_B(\vec\eta',{\bf -k}))}{|{\bf k}|^2} 
 \left[ \frac{\eta'.k_\flat}{\eta.k_\flat}\right]_{k_{\flat} = (s |{\bf k}|,\vec{\bf k})}
\eea
where $s=\sign ({\bf k}.(\vec\eta-\vec\eta'))$, $\eta \equiv (r,\vec\eta)$, and $\delta_{\perp}^{(2)}$ is a delta-function in the plane transverse to $k$.
This is the analogue of the ``Coulomb term'' $\frac{1}{2} J^0 \frac{1}{\grad^2} J^0$ in the Hamiltonian for Coulomb-gauge QED. 
In the second line, we've presented it in a form that is manifestly projective in $\eta$-space (for $\kappa=0$). 
In momentum 3-space, the interaction energy scales as $\sim 1/|{\bf k}|^2$, so we expect the potential to fall off
at least as fast as $\sim1/R$ in position space. The presence of the phase factor when $\kappa\neq 0$ will 
if anything increase the fall-off at large $R$ (soft-$|{\bf k}|$). 

To understand how the physics is being encoded, we work through a couple of examples, taking $\kappa=0$,
and using the projective version of these results. The only thing 
that changes is that the measures are replaced by their projective counterparts, and the $\eta_\perp^2 \delta_{\perp}^{(2)}$
is replaced by the angular part of the delta function $\delta(\varphi(\vec\eta_{\perp})-\varphi(\vec\eta'_{\perp}))$.
The interaction energy is then
\be
V_{static} = \int d^3{\bf k} \frac{d^3\vec\eta\delta(g-1)}{4r^3} \frac{d^3\vec\eta'\delta(g'-1)}{4r'^3} 
\delta(\varphi(\vec\eta'_{\perp})-\varphi(\vec\eta_{\perp})) \frac{(r \tilde{J}_B(\vec\eta,{\bf k})) (r' \tilde{J}_B(\vec\eta',{\bf -k}))}{|{\bf  k}|^2} 
 \left[ \frac{\eta'.k_\flat}{\eta.k_\flat}\right]. \label{eq:example1}
\ee

The simplest example (and the only one we have so far made manifestly covariant and local) is to introduce a current $J(\eta,x)=J(x)$ independent of $\eta$, which trivially satisfies the $\kappa=0$ continuity condition.  In this case $J_B=0$ so the Coulomb term vanishes, while $J_A(\vec\eta,x)=J(x)$.
At the level of the covariant projective action \eqref{eq:BaseAction}, the interaction term is
\be
\int d^4x J(x) d\mu(\eta) \delta'(\eta^2)\psi(\eta,x) = \int d^4x J(x) \psi(x)
\ee
where $\psi(x)$ is the gauge invariant ``scalar'' component
\be
\psi(x) = \int d\mu(\eta) \delta'(\eta^2)\psi(\eta,x).
\ee
This ``scalar'' projection is only local and gauge invariant for $\kappa=0$. 
Thus, this interaction describes a standard looking matter sector coupled to a single scalar component of $\psi$.

Several new subtleties arise in coupling the homogeneous-action fields to vector currents --- first, there is no natural covariant and local mapping from a vector $J^\mu(x)$ to a homogeneous field $J(\eta,x)$.   We can, however, embed a current $J^\mu(x)$ into $J(\eta,x)$ in a manner that is \emph{not} manifestly covariant, but still recovers the physics of a gauge boson coupled to a vector current.  The simplest such coupling is 
\bea
\tilde{J}_A(\vec\eta,{\bf k}) &=& \frac{1}{|{\vec \eta_\perp}|} \vec\eta \cdot \tilde{\vec J}({\bf k}), \\
\tilde{J}_B(\vec\eta,{\bf k}) &=& - \frac{|{\vec \eta}|}{|{\vec\eta_\perp}|}\tilde J^0({\bf k})\label{eq:JAB}
\eea
where $|\vec\eta_\perp|^2 = \vec\eta^2 - (\eta.\vec{\bf k})^2/|{\bf k}|^2$ and the components $\tilde J^\mu({\bf k})$ satisfy  the usual conservation condition 
$p\cdot J(p)=0$.  This form does not satisfy the requirements to obtain a bounded action (bounded solutions to the continuity condition involve more complex dependence of $J_A$ and $J_B$ on $J_0$ and $\vec J$), but already gives some intuition for how the ordinary gauge-theory structure is encoded.

In this case the factor of $J_0^2/|{\bf k}|^2$ can be factored out of the $\vec\eta$ and $\vec\eta'$ integrals in \eqref{eq:example1}.  The angular integrals are trivial, while the $z$ integral must be regulated becase of the unboundedness of our currents.  Regardless, the static term takes the form 
\be
V_{static} \propto \int d^3{\bf k} \frac{J^0({\bf k})J^0({\bf -k})}{|{\bf k}|^2}.
\ee
which is precisely the structure of the QED Coulomb term. 
We can also write the $J_A$ coupling as,
\bea
-\int d^3 k \frac{d^3\vec\eta\delta(g-1)}{4 r^3} \tilde{J}_A(\vec\eta,{\bf k}) \tilde{A}(\vec\eta,-{\bf k}) 
&=& - \int d^3 k \tilde{\vec J}({\bf k}) \cdot \tilde{\vec A}(-{\bf k}) \nonumber \\
&=& -\int d^3 x \vec{J}(x)\cdot \vec{A}(x) \label{eq:vectorcoupling}
\eea
where 
\be
\tilde{\vec A}({\bf k}) 
= \int\frac{d^3\vec\eta\delta(g-1)}{4 r^3} |{\vec \eta}| \grad_\eta A(\vec\eta,{\bf k})
= \int\frac{d^3\vec\eta\delta(g-1)}{4 r^3} \frac{\vec \eta_\perp({\bf k})}{|\vec\eta_\perp({\bf k})|} A(\vec\eta,{\bf k})
\ee
is the ``vector'' component of $A(\eta,{\bf k})$, which by construction satisfies the Coulomb gauge condition $\grad_x.\vec A(x)=0$. 
Thus, the current $J_A$ and $J_B$ given above reproduces (up to constant coefficients that we have not computed) quantum electrodynamics in Coulomb gauge. 
A more thorough discussion will be given in \cite{ST:inprep}.
One key lesson is that single tensor-like objects $J^{\mu}(x), A^{\mu}(x)$, etc.~ are difficult to extract in a manifestly local and covariant manner
from $J(\eta,x)$ and $\psi(\eta,x)$. 
It is not clear whether the difficulty of coupling local currents to CSPs is qualitatively different from the difficulty encountered here. 

%%%%%%
\section{Particle Degrees of Freedom}\label{dof}

In $\deltabar A =0 $ gauge, independent modes of $A$ are labelled by functions of the \emph{transverse} components of $\eta$, i.e. $\vec\eta_\perp$ 
and are therefore a two-parameter family.  We now wish to understand the physical interpretation of this family.

As the theory is Lorentz invariant, the correct way to identify the particle degrees of freedom is to construct 
states that are simultaneous eigenstates of $P^{\mu}$ and the Pauli-Lubanski invariant $W^2$, where  
\be
W^\mu = \frac{1}{2} \epsilon^{\mu\nu\rho\sigma}J_{\nu\rho} P_{\sigma}.
\ee
In Appendix \ref{app:poincare}, we construct canonical momentum and Lorentz generators and calculate their commutators with the canonical fields $A$ and $\Pi_\perp$ --- we'll use those results to carry out the above diagonalization, and then express the field operators in terms
of these eigenstates. 
Since our free fields satisfy $\Box A = 0$, their excitations are massless particles.  
We will see that $W^2$ only annihilates the fields for $\kappa=0$, implying that for $\kappa \neq 0$ our action propagates 
continuous-spin particles (CSPs).  CSPs are characterized by the non-zero eigenvalue $W^2 \ket{\psi}= -\rho^2 \ket{\psi}$, 
and our action propagates CSPs with \emph{all} $\rho$'s.  The Hamiltonian can naturally be decomposed into ``sectors'' with different 
$\rho$ that are separately Lorentz-invariant.  The detailed form of the current $J$ will single out some combination of CSPs as interacting, while others are free.  We comment in \S\ref{singleCSP} on a gauge-fixed Hamiltonian for a single CSP, though we do not know of an action from which it can be derived.  

We begin by summarizing some relevant facts about the Lorentz generators and their action on fields.
In $\deltabar A =0$ gauge, all generators take a form very similar to the generators in a free scalar theory, except for the $\vec\eta\cross\grad_\eta$ term in ${\bf J}$, which ensures that ${\bf J}$ generates rotations in $\eta$-space as well as $x$-space.   The Lorentz commutators with the fields
\bea
&&[{\bf J}^{i},A(\vec \eta,{\bf x})] = -i ({\bf x}\cross \grad_x + \vec \eta \cross \grad_\eta)^i A(\vec\eta,{\bf x}) \\
&&[{\bf J}^{i},\Pi_\perp(\vec \eta,{\bf x})] = -i ({\bf x}\cross \grad_x + \vec \eta \cross \grad_\eta)^i \Pi_{\perp}(\vec\eta,{\bf x}) \\
&&[{\bf K}^i,A(\vec \eta,{\bf x})] = -i x^0 \nabla_i A(\vec \eta,{\bf x})  + i \left(x^i -  \f{\eta.\grad_x \grad_\eta^i}{\grad^2_x}\right) \Pi_{\perp}(\vec\eta,{\bf x}) \label{KAtext}\\
&&[{\bf K}^i,\Pi_\perp(\vec \eta,{\bf x})] = -i x^0 \nabla_i \Pi_\perp(\vec \eta,{\bf x})  + i \left(x^i \grad^2_x -  \eta.\grad_x \grad_\eta^i\right) A(\vec\eta,{\bf x}).\label{KPitext}
\eea
 are each annihilated by $\deltabar$ by construction. The non-local term in the action of boosts ${\bf K}^i$ on $A$, elaborated on in Appendix \ref{app:poincare}, is reminiscent of Coulomb-gauge QED and is needed to ensure that the boosted field remains in $\deltabar A = 0$ gauge.

Now, let us consider operators that are eigenstates of $P^{\mu}$
\be
\Psi_-({\bf p},\vec\eta)\equiv \int d^3x \left[|{\bf p}|A({\bf x},\vec\eta) - i\Pi_\perp({\bf x},\vec\eta)\right] e^{-i{\bf p}.{\bf x}}.
\ee
Using 
\begin{align}
[H,A(x,\vec\eta)]&=-i\Pi_{\perp}(x,\vec\eta) &\quad [H,\Pi_{\perp}(x,\vec\eta)]& =-i\nabla_x^2A(x,\vec\eta) \\
[P^i,A(x,\vec\eta)]&=-i\nabla_x^iA(x,\vec\eta)& \quad [P^i,\Pi_{\perp}(x,\vec\eta)]& =-i\nabla_x^i\Pi_{\perp}(x,\vec\eta)
\end{align}
it is easy to check that these operators have eigenvalues $p^{\mu} \equiv (|{\bf p}|,{\bf p})$.  
Choosing polarization ``vectors'' $\epsilon_\pm^\mu({\bf p})$ with 
\be
\epsilon_- = \epsilon_+^* \quad \epsilon_\pm.p_\pm = 0 \quad \epsilon_+.\epsilon_- = -1 \quad \epsilon_+^2 = \epsilon_-^2 = 0 \quad \epsilon_\pm^0 = 0 \quad \vec\epsilon_+ \cross \vec\epsilon_- = -i \hat {\bf p}.
\ee
we can decompose $W^2 = - W_+ W_-$ where 
\be
W_\pm = W.\epsilon_\pm  = {\vec\epsilon}_\pm .({\bf K}\cross {\bf p} +|{\bf p}| {\bf J}).
\ee
so that
\bea
[W_\pm, \Psi({\bf p},\vec\eta) ] = -i |{\bf p}| {\vec\epsilon}_\pm . ({\vec \eta}_\perp \cross \grad_\eta)  \Psi({\bf p},\vec\eta),
\eea
where $\vec \eta_\perp  \equiv \vec\eta - \frac{\vec\eta.{\bf p}}{|{\bf p}|^2} {\bf p}$.

%%%%%%%%
\subsection{$W_\pm$ eigenstates}

Since $[W_+,W_-]$ commute they can be simultaneously diagonalized. $(W_+)^{\dagger}=W_-$, so the eigenvalues 
are complex conjugates of one another and we can seek out functions $f^{+}$ such that
\be
a^\dagger_{p,\rho,\phi} \equiv \int \frac{d^3\eta}{4 r^3} \Psi_-({\bf p},\vec\eta) f^+_{\rho,\phi}(\vec\eta,{\bf p}).
\ee
diagonalizes
\be
W_\pm a^\dagger_{p,\rho,\phi} = \rho e^{\pm i\phi} a^\dagger_{p,\rho,\phi}.
\ee
These $f$'s must therefore satisfy 
\be
-i |{\bf p}| {\vec\epsilon_\pm}.(\vec\eta_\perp \cross \grad_\eta) f^+_{\rho,\phi}(\vec\eta,p) = 
(\vec\eta_\perp\cross \vec\epsilon_\pm).\grad_\eta f^+_{\rho,\phi}(\vec\eta,p) = i \frac{\rho}{|{\bf p}|} e^{\pm i\phi} f^+_{\rho,\phi}(\vec\eta,p),\label{Weqn}
\ee
in addition to our gauge condition $\deltabar f = 0$.  
Because $\vec\eta_\perp$ and $\vec\epsilon$ are both in the plane transverse to ${\bf p}$, the differential operator in \eqref{Weqn}, 
like $\deltabar$, is proportional to ${\bf p}.\grad_{\eta}$.
Defining 
\be
\epsilon(p\phi) \equiv \frac{i}{\sqrt{2}}( \epsilon_{+}e^{-i\phi}-\epsilon_{-}e^{i\phi}) = \sqrt{2} i \Im[\epsilon_+ \, e^{-i\phi}] \quad \mbox{with} \quad \epsilon(p\phi)\cross \epsilon_\pm = \frac{1}{\sqrt{2}}e^{\pm i\phi} \hat k,
\ee
we find that the unique simultaneous solution to these two equations (for $\kappa \neq 0$) is
\be
f^+_{\rho,\phi}(\vec\eta,p) \equiv \frac{2\rho^2|{\bf p}|}{\kappa^2}  \int d\tau e^{i\tau \kappa} \delta^3\left(\vec\eta - \tfrac{\sqrt{2} \rho}{\kappa}\vec\epsilon(p,\phi) - \tau \vec p\right),\label{fsolution}
\ee
where the pre-factor is chosen for later convenience. 
By Fourier-transforming these we can obtain
\be
f^+_{\rho,n}(\vec\eta,p) \equiv \int \frac{d\phi}{2\pi} e^{in\phi} f^+_{\rho,\phi}(\vec\eta,p) 
\ee
for all integer $n$, which diagonalize the rotation generator $\hat {\bf p}.{\bf J}$.  
Generic Lorentz transformations will mix the different $\phi$ (or $n$) eigenstates, but leave $\rho$ invariant.  
In the case $\kappa = 0$, neither $\deltabar f = 0$ nor \eqref{Weqn} restricts the profile of $f$ in $\eta_\perp$.  
Solutions to \eqref{Weqn} are therefore non-unique.  Had we used the homogeneous action for $\kappa=0$, the homogeneity requirement $\vec\eta.\grad_\eta f = 0$ would single out the profile 
\be
f^+_{\phi}(\vec\eta,p) \equiv |{\bf p}| \int dr d^3 d\tau \delta^3\left(\vec\eta -  r \vec\epsilon(p,\phi) - r \tau \vec p\right).
\ee
These can again be transformed to Fourier-conjugate helicity basis states, which in the case $\kappa=0$ are Lorentz-invariant.

The mode expansion for the field $A(\vec\eta,x)$ (in the interaction picture) is now
\be
A(\vec\eta,x) = \int \frac{d\rho}{\rho} \frac{d^3{\bf p}}{2|{\bf p|}}\frac{d\phi}{2\pi} \left( e^{-i p\cdot x}f^+_{\rho,\phi}(\vec\eta,{\bf p})a_{{\bf p},\rho,\phi}+e^{+i p\cdot x}f^-_{\rho,\phi}(\vec\eta,{\bf p})a^{\dagger}_{{\bf p},\rho,\phi} \right).
\ee
and we can easily verify that the field commutation relation are equivalent to
\be
[a_{{\bf p},\rho,\phi},a^{\dagger}_{{\bf p'},\rho',\phi'}] = 2\pi \delta(\phi-\phi')\rho\delta(\rho-\rho')2|{\bf p}| \delta^{3}({\bf p}-{\bf p'}).
\ee
The other commutators $[a_{{\bf p},\rho,\phi},a_{{\bf p'},\rho',\phi'}]$ and $[a^{\dagger}_{{\bf p},\rho,\phi},a^{\dagger}_{{\bf p'},\rho',\phi'}]$ vanish. 
This is precisely the Lorentz invariant commutator algebra we'd expect for CSP states!
As one would expect, the free part of the Hamiltonian can be expressed in terms of the annihilation and creation operators
as
\be
{\bf H}_{free} = \int \frac{d^3{\bf p}}{2|{\bf p|}}\frac{d\rho}{\rho}\frac{d\phi}{2\pi} \left( |{\bf p}|a^{\dagger}_{{\bf p},\rho,\phi}a_{{\bf p},\rho,\phi} \right),
\ee
which further confirms that we have a unitary theory of CSPs for all $\rho\geq 0$.

To finish the analysis, we expand about a vacuum $\ket{0}$ annihilated by all mode operators $a({\bf p},\rho\phi)$.
We then define familiar particle states,
\be
\ket{{\bf p}\phi}_{\rho} \equiv a^{\dagger}_{{\bf p},\rho,\phi}\ket{0}.
\ee
Under Lorentz transformations,
$
(\Lambda^{-1})^{\mu}_{\nu}\epsilon^{\nu}(\Lambda p, \phi+\theta) = \epsilon^{\mu}(p,\phi) - \sqrt{2}Re( \beta e^{-i(\phi+\theta)})p^{\mu},
$
where $\theta(\Lambda,p)$ and $\beta(\Lambda,p)$ have the interpretation of E(2) Little Group rotations and translations (see e.g. \cite{Schuster:2013pxj}).
Using this relation, and using our expressions for the Lorentz generators from Appendix \ref{app:poincare}, we can readily compute
\be
U(\Lambda)\ket{{\bf p}\phi}_{\rho} = e^{i\rho Re(\sqrt{2}\beta e^{-i(\phi+\theta)})}\ket{\Lambda {\bf p}(\phi+\theta)}_{\rho}
\ee
The phase factor is just the little group E(2) translation factor of the continuous spin representation with 
spin-scale $\rho$ \cite{Schuster:2013pxj}. Thus, this theory is a quantum theory of CSP particles with spin-scale $\rho$. The Hilbert
space of this theory contains all CSPs with $\rho\geq 0$. States with different $\rho$ do not mix under Lorentz transformations. 

\subsection{Connection to Covariant CSP Wavefunctions}
The wavefunctions \eqref{fsolution} are \emph{not} covariant, in precisely the same sense that photon polarization tensors $\epsilon^\mu$ are non-covariant: Little Group transformations on the state labels $\rho$ and $\phi$ are equivalent to the action of the canonical Lorentz generators ${\bf J}$ and ${\bf K}$.  But the action of these generators is not a simple Lorentz-transformation $x\rightarrow \Lambda^{-1} x$, $\eta \rightarrow \Lambda^{-1} \eta$.  Rather, they have a non-covariant term coming from the $\vec\eta.\grad_x \grad_\eta^i$ piece of the commutation relations \eqref{KAtext} and \eqref{KPitext}, which can be interpreted as precisely the gauge transformation that brings a Lorentz-transformed field back to $\deltabar A=0$ gauge.  Thus Lorentz or Little Group actions on the canonical field are equivalent to Lorentz transformations of $\eta$ and $x$ \emph{only up to a gauge transformation}, like gauge-fixed fields in QED.

Past efforts to construct field theories for CSPs \cite{Yngvason:1970fy,Iverson:1971hq,Chakrabarti:1971rz,Abbott:1976bb,Hirata:1977ss} have all constructed fields assuming that they have a mode expansion in terms of  \eqref{covariant} or similarly covariant wavefunctions.  That these studies encountered non-vanishing commutators at spacelike separations is hardly surprising.  The wavefunctions \eqref{covariant} are localized on $\delta(\eta.p)$ --- it is difficult to envision how this singular momentum-dependent support could ever arise from a local field theory!  The results of these past studies contrast sharply with our theory, whose fields \emph{do} commute at equal times.  

There is, however, a simple mathematical relationship between our $f$'s and one of the families of covariant wavefunctions classified in \cite{Schuster:2013pxj}, namely
\be
\psi_{p,\phi\rho}(\eta^\mu) = \int d\tau e^{i\tau\kappa} \delta^{(4)}(\eta -\sqrt{2} \frac{\rho}{\kappa}\epsilon(p\phi) - \tau p).\label{covariant}
\ee
Our wavefunctions can be simply obtained by integrating these covariant $\psi$'s over $\eta^0$: 
\be
f_{p,\phi\rho}(\vec\eta) = \int_{-\infty}^{\infty} d\eta^0 \psi_{p,\phi\rho}(\eta^\mu).
\ee
Though this clearly requires a Lorentz-breaking choice (of the $\eta^0$ direction along which to integrate), it removes the unnatural localization on $\delta(\eta.p)$ and allows for well-behaved equal time commutators of fields.  We have not yet found a physical interpretation of this integration trick, but it might shed light on generalizations of the action \eqref{eq:BaseAction}, or on its proper interpretation.

%%%%%%%%%%%%%%%%%%%%%%%%%%%
\section{Projective Hamiltonian Field-Theory for a Single CSP}\label{singleCSP}

The local action that we used to derive our CSP field theory contains an entire E(2) plane worth of 
particle states. While this plane has a clear phyiscal interpretation, it is natural to try to build a theory for a single CSP with fixed $\rho$, as states with different $\rho$ do not mix under Lorentz transformations. 
One obvious and simple way to do this is to render the action \eqref{eq:BaseAction} projective, as discussed 
in Section \ref{sec:action} and Appendix \ref{app:homog}.  It is not clear how to extend this projective trick to $\kappa \neq 0$ in a manifestly local and covariant way.
Nonetheless, it is trivial to take our final form for the gauge-fixed Hamiltonian CSP field theory, and make this theory projective! 
 This is achieved by sending $\kappa\rightarrow \rho/|\vec\eta_{\perp}(p)|$ in the gauge-fixing condition and in all of our final results, 
 accompanied by a replacement of the measure $\frac{d^3\eta}{4r^3}$ with the projective measure $D^2\eta \equiv \frac{d^3\eta}{4r^3}\delta(g-1)$
 with $g$ a homogenous function of $\eta$. 

This procedure is implemented at the level of the $\deltabar$-gauge fixing and quantization conditions
used to define the theory. The modified gauge conditions are easiest to state in momentum space. 
The modified gauge fixing condition, obtained by $\kappa\rightarrow \rho/|\vec\eta_{\perp}(p)|$ is
\be
\deltabar_P A(\vec\eta,\vec p) = \deltabar_P \Pi(\vec\eta,\vec p) = 0 \qquad \mbox{ with } \deltabar_P \equiv \left(-i|\vec\eta_{\perp}(\vec p)|\vec p \cdot \nabla_{\eta} +\rho \right). 
\ee
These conditions are  compatible with the homogeneity requirements
\be
\vec\eta\cdot \vec\nabla_{\eta}A=\vec\eta\cdot \vec\nabla_{\eta}\Pi=0,
\ee
which implies that the fields are functions of rays in $\vec\eta$-space, not points. 
On these constraints, the theory is defined by the equal-time commutators of the fields
\bea
&&[A(\eta,{\bf x}),A(\eta',{\bf x'})] = [\Pi(\eta,{\bf x}),\Pi(\eta',{\bf x'})] = 0 \nonumber \\
&&[A(\eta,{\bf x}),\Pi(\eta',{\bf x'})] = i\bar\delta_P({\bf x}-{\bf x}',\eta , \eta' ),
\eea
where
\bea
\bar\delta_P({\bf x}-{\bf x}',\eta , \eta' ) &\equiv& \int d^3 {\bf p} e^{i{\bf p}\cdot ({\bf x}-{\bf x'})} \tilde{\delta}_P({\bf p},\eta,\eta') \nonumber \\
\tilde{\delta}_P({\bf p},\eta,\eta') &=& \delta(\varphi(\vec\eta) - \varphi(\vec\eta')) \times \exp \left[-i \rho \left(\frac{\eta_z}{\eta_\perp} - \frac{\eta'_z}{\eta'_\perp}\right)\right].
\eea
and
$\varphi(\eta)$ is the angular coordinate of $\vec \eta$ in the $\hat p$-centric coordinates. 

The form of these commutators is essentially dictated by consistency with the constraints --- $\bar\delta_P$ projects generic functions of $\eta,x$ onto functions satisfying the constraints. In particular, for any (local) operator ${\mathcal O}(x)$ built out of $A$ and $\Pi$, the structure of $\bar\delta_P$ implies
\be
[\deltabar_P \Pi(\eta,{\bf x}), {\mathcal O}(x)] = 0
\ee
and similarly for $A$.
Thus, the constraint equations are consistent with the canonical quantization prescription. 
On the space of operators $\mathcal{O}(\eta,x)$ satisfying the constraints, 
\be
\int D^2\eta' d^3y \bar\delta_P(x-y,\eta,\eta')\mathcal{O}(y,\eta') = \mathcal{O}(x,\eta),
\ee
so that $\bar\delta_P(x,\eta,\eta')$ behaves like a delta function on the support of the gauge condition.

The free Hamiltonian 
\be
{\bf H} = \int d^3{\bf x} D^2\eta \f{1}{2} \left( \Pi(\eta,{\bf x})^2 + |\grad A({\eta},{\bf x})|^2 \right),
\ee
where $D^2\eta=\frac{d^3\eta}{4r^3}\delta(g-1)$, is local, as are the other \Poincare generators which are identical to those of Appendix \ref{app:poincare} except that the integrals now use the projective $D^2\eta$ measure.  
The equations of motion are
\bea
\partial_{t}A(\eta,{\bf x}) &\equiv& i[{\bf H}, A(\eta,{\bf x})] \\
\partial_{t}\Pi(\eta,{\bf x}) &\equiv& i[{\bf H}, \Pi(\eta,{\bf x})] ,
\eea
so that
\be
\partial^{\mu}\partial_{\mu} A(\eta,x) = 0.
\ee

The Lorentz algebra
and compatibility of Lorentz transformations with the gauge and projective conditions
works as it did before. 

We identify the particle degrees of freedom by expanding our new field operators 
in terms of operators that simultaneously diagonalize $P^{\mu}$ and $W^2$.
In this case, the appropriate operators are
\bea
a({\bf p},\phi) &\equiv& \int D^2\eta d^3x e^{-i{\bf p}\cdot {\bf x}} F^{-}(\eta,{\bf p}\phi)\left( |{\bf p|}A(\eta,x)+i\Pi(\eta,x) \right)_{x^0=0} \\
a^{\dagger}({\bf p},\phi) &\equiv& \int D^2\eta d^3x e^{+i{\bf p}\cdot {\bf x}} F^{+}(\eta,{\bf p}\phi)\left( |{\bf p|}A(\eta,x)-i\Pi(\eta,x) \right)_{x^0=0} 
\eea
where
\be
F^{\pm}(\vec{\eta},{\bf p}\phi) = \int_{r>0} dr d\tau |{\bf p}|r^2 (2\pi) \delta^3(\vec{\eta}-r\vec{\epsilon}({\bf p},\phi))-r\tau {\bf p})e^{\mp i\rho\tau}.
\ee 
It's straightforward to verify that the field commutation relations are equivalent to
\be
[a({\bf p},\phi),a^{\dagger}({\bf p'},\phi')] = 2\pi \delta(\phi-\phi')2|{\bf p}| \delta^{3}({\bf p}-{\bf p'})
\ee
which shows that these modes act like annihilation and creation operators for a single CSP.
Computing $W_+$ and $W_-$ as we did before (with the projective substitutions), it's easy to check that
\be
W^2a^{\dagger}({\bf p},\phi) \ket{0} = -W_+W_- a^{\dagger}({\bf p},\phi) \ket{0} = -\rho^2 a^{\dagger}({\bf p},\phi) \ket{0}
\ee
which show that we have a single CSP with spin-scale $\rho$.
The Hamiltonian expressed in terms of annihilation and creation operators is 
 \be
{\bf H} = \int \frac{d^3{\bf p}}{2|{\bf p|}}\frac{d\phi}{2\pi} \left( |{\bf p}|a^{\dagger}({\bf p},\phi)a({\bf p},\phi) \right),
\ee
and the mode expansion for the field $A(\eta,x)$ at an arbitrary time is just,
\bea
A(\eta,x)= && e^{-i{\bf H}x^0}A(\eta,{\bf x})e^{+i{\bf H}x^0} = \nonumber \\
&&\int \frac{d^3{\bf p}}{2|{\bf p|}}\frac{d\phi}{2\pi} \left( e^{-ip\cdot x}F^+(\eta,{\bf p}\phi)a({\bf p}\phi)+e^{+ip\cdot x}F^-(\eta,{\bf p}\phi)a^{\dagger}({\bf p}\phi) \right).
\eea
Defining the particle state as we did before, we find that
\be
U(\Lambda)\ket{{\bf p}\phi} = e^{i\rho Re(\sqrt{2}\beta e^{-i(\phi+\theta)})}\ket{\Lambda {\bf p}(\phi+\theta)}
\ee
The phase factor is just the little group E(2) translation factor of the continuous spin representation with 
spin-scale $\rho$. Thus, this theory is a quantum theory of CSP particles with spin-scale $\rho$.  
While the gauge-fixing condition is spatially non-local, the Hamiltonian and all other canonical \Poincare generators for this theory are local in the fields.  

%%%%%%%%%%%%%%%%%%%%%%%%%%%%%%%%%%%%%%%%%%%%
\section{Conclusions and Future Directions}\label{sec:conclusion}

We have proposed a local and covariant action \eqref{eq:BaseAction} that defines a classical gauge theory of CSPs. We have quantized this theory in the presence 
of background currents, which were assumed to satisfy the continuity condition \eqref{eq:Continuity} needed for gauge invariance and to commute with themselves at equal times.
The local degrees of freedom are described by a field that depends on both a space-time coordinate $x^\mu$ and a spin-space four-vector $\eta^{\mu}$. 
%This description allows us to easily unify massless helicity, continuous-spin, and massive spin degrees of freedom, 
%and generalizes to any number of dimensions.
This description allows us to easily unify massless helicity and continuous-spin degrees of freedom, 
and generalizes to any number of dimensions.
 The action is localized to the neighborhood of the positive null-$\eta$ cone; in $D$ dimensions a linear gauge invariance reduces the physical 
information to a single function on a Euclidean $(D-2)$-plane, on which the little group $E(D-2)$ acts naturally.  
A projective version of the action further reduces the physical content to $S^{D-3}$, enabling a new description of particles 
with any spin type.
Our theory is the first quantum field theory of any kind for continuous spin particles that admits a description in terms of local fields. 

All previous attempts to build a quantum theory started with Wigner's original formulation of CSPs in terms of wavefunctions with {\it singular} 
momentum space support, or other covariant fields without a gauge redundancy. 
Such a formulation fails to yield a consistent quantum theory of helicity 1 or 2, so it is not surprising that it also fails for CSPs! 
Our hope is that the existence of a healthy classical and quantum theory of CSPs, as formulated here,
will renew a proper investigation of these theories.

Among the reasons to seek a theory of interacting CSPs are the existence of covariant soft factors \cite{Schuster:2013pxj} (in contrast to high-spin massless particles) and the correspondence of these soft factors and ansatz amplitudes to those of scalars, photons, and gravitons at energies much larger than $\rho$ \cite{Schuster:2}.   These two results suggest that CSP interactions may be both consistent and relevant to Nature!  
These $S$-matrix results also exhibited potentially problematic and apparently non-local features.  
Foremost among these is the dependence of the soft factors of \cite{Schuster:2013pxj} on a momentum-dependent phase factor.  
While this phase is everywhere bounded and presents no immediate physical inconsistency, 
it does cast doubt on whether the soft factors can be obtained from a local theory.  
The deformation of the continuity condition $\delta(\eta^2)(\partial_x\cdot\partial_{\eta})J = 0$ to the \emph{local} \ $\delta(\eta^2)(\partial_x\cdot\partial_{\eta} +\kappa)J = 0$ 
condition seems to be what gives rise to the apparently {\it non-local} phase structure in the soft-factors of \cite{Schuster:2013pxj}. 
Our first attempts to build local currents $J(\eta,x)$ satisfying \eqref{eq:Continuity} focused on $J$'s constructed from a single $\eta$-independent matter field $\phi(x)$, and in this simple case we have not found any local theory that works. 
It is interesting that a local continuity condition is sufficient to furnish linear interactions, but 
a local solution to this condition may require additional structure or symmetry. 
While our formalism has allowed us to frame the problem of interactions sharply, we have only started to make  progress towards settling these questions. 

Several directions appear promising towards modeling interactions.  One possibility is that currents are more readily built out of matter fields that are themselves functions of $\eta$.  In addition, there appear to be generalizations of the action \eqref{eq:BaseAction} that involve higher $\eta$-derivatives but still two spatial derivatives.  Classifying these and establishing their relationship to our original action is important. 
The action \eqref{eq:BaseAction} appears most naturally suited to describing the interactions of the ``scalar'' component of $\psi(\eta,x)$,
and it may be that related (but equivalent) actions are better suited for describing ``vector'' and ``graviton'' components. 
A related question is whether the single-CSP Hamiltonian theory of Section \ref{singleCSP} can be obtained from some covariant action, perhaps by generalizing the projectivisation trick used for $\kappa=0$.  
In a different direction, it would be interesting to find generalizations of this action to curved space, and to see whether they can make contact with Vasiliev's theory in AdS backgrounds\cite{Vasiliev:1988xc,Vasiliev:1988sa,Vasiliev:2004qz}. Connections to other descriptions of high-spin degrees of freedom would also be interesting to explore \cite{Sorokin:2004ie,Bouatta:2004kk,Bekaert:2005vh,Fotopoulos:2008ka,Benincasa:2011pg}.

At the same time, important aspects of even our simple free field theory \eqref{eq:BaseAction} remain obscure.  A major gap in our present understanding of this theory is a classification of local gauge-invariant operators.  Although the propagating degrees of freedom of \eqref{eq:BaseAction} are the same as those of tensor gauge theories, the packaging of these degrees of freedom in \eqref{eq:BaseAction} appears to be quite different.  It is not clear, for example, whether four-vector gauge fields $A^\mu$ are encoded covariantly in $\psi$, and so it is far from obvious if and how something like a ``field strength'' $F^{\mu\nu}$ can be obtained.  The conserved Belinfante stress-energy tensor is only gauge-invariant under \eqref{gaugePsiA} up to a total derivative term, which is suggestive of a non-linear gauge invariance analogous to that of helicity-2 gravity. These and other questions warrant further investigation.  

%%%%%%%%%%%%%%%%%%%%%%%%
\section*{Acknowledgments}
We thank 
Haipeng An,
Nima Arkani-Hamed,
Cliff Burgess, 
Freddy Cachazo,
Lenny Evans,
Jared Kaplan,
Ami Katz,
Yasunori Nomura,
Maxim Pospelov, 
Yanwen Shang,
Carlos Tamarit,
Jesse Thaler,
Mark Wise,
and Itay Yavin
for helpful discussions on various aspects of this physics and Carlos Tamarit for feedback on the manuscript. 
We especially thank Jin-Mann Wang for identifying several mistakes in an earlier version of section 2.  
Research at the Perimeter Institute is supported in part by the Government of Canada through NSERC and by the Province of Ontario through MEDT. 

%%%%%%%%%%%%%%%%%%%%%%%%%%%%%%%%%%%%%%%%%%%%
%%%%%%%%%%%%%%%%%%%%%%%%%%%%%%%%%%%%%%%%%%%%
%%%%%%%%%%%%%%%%%%%%%%%%%%%%%%%%%%%%%%%%%%%%
%%%%%%%%%%%%%%%%%%%%%%%%%%%%%%%%%%%%%%%%%%%%
%%%%%%%%%%%%%%%%%%%%%%%%%%%%%%%%%%%%%%%%%%%%
%%%%%%%%%%%%%%%%%%%%%%%%%%%%%%%%%%%%%%%%%%%%
%%%%%%%%%%%%%%%%%%%%%%%%%%%%%%%%%%%%%%%%%%%%
%%%%%%%%%%%%%%%%%%%%%%%%%%%%%%%%%%%%%%%%%%%%
%%%%%%%%%%%%%%%%%%%%%%%%%%%%%%%%%%%%%%%%%%%%
\appendix

%%%%%%%%%%%%%%%%%

\section{Conventions for Homogeneous Action and Hamiltonian}\label{app:homog}
For the homogeneous action, we use the homogeneity $4$ measure
\be
d\mu(\eta) \equiv d^4\eta \delta(g(\eta)-1) 
\ee
where $g(\eta)$ is an arbitrary homogeneity-1 function of $\eta$ corresponding to the choice of ``representative'' $\eta$ on each ray.  This has the property that
\be
d\mu(\eta) \equiv d^4\eta \delta(g(\eta)-1) f^{(-4)}(\eta)
\ee
is independent of the choice of $g$ for any $f$ of homogeneity $-4$, i.e. $\eta.\partial_\eta f = -4$.  
There is no weight-zero Lorentz-invariant measure. 

When we break Lorentz-invariance in the $A-B$ decomposition or in building the Hamiltonian, we will switch to a weight-zero measure 
\be
D^2 \vec\eta \equiv \frac{d^3\vec\eta}{4 |\vec\eta|^3} \delta(g(\vec\eta)-1).
\ee

A projective $\delta$-function that works nicely with this measure is 
\be
\delta_P(\vec\eta,\vec\eta') \equiv \int_{0,\infty} \frac{da}{a} |\vec\eta|^3 \delta^3(\eta-a \eta') = \int_{0,\infty} \frac{da}{a} |\vec\eta'|^3 \delta^3(\eta a - \eta').
\ee

For a field $X$ of homogeneity $h$, we define functional derivatives as
\be
\frac{\delta A(\vec\eta,\vec x)}{\delta A(\vec\eta',\vec x')} \equiv \left(\frac{|\vec\eta|}{|\vec\eta'|}\right)^h \delta^3(\vec x - \vec x')  \delta_P(\vec\eta,\vec\eta').
\ee
This produces the intuitive result that
\be
\frac{\delta}{\delta A(\vec\eta',\vec x')} \int d^3 x D^2\vec\eta A(\eta,x) B(\eta,x) = B(\eta',x').
\ee
It is important, however, to account for the factor of $1/|\vec\eta|^3$ in $D^2\eta$ when integrating by parts, so that 
\be
\int D^2\eta f(\vec\eta)\grad_\eta^i g(\vec\eta) =  - \int D^2\eta r^3 \grad_\eta^i \left(\tfrac{1}{r^3} f(\vec\eta)\right) g(\vec\eta).
\ee

\section{An $\vec\eta$-Space Green's Function}\label{app:greens}
An operator that recurs several times in the classical and quantum theories is
\be
D \equiv r \deltabar r^3 \deltabar \tfrac{1}{r^2}.
\ee
This section defines a Green's function for $D$, solves for it explicitly, and summarizes some important properties. 
We introduce a Green's function defined by 
\be
D G_D(\vec\eta,\vec\eta',x,x') = \bar\delta(\vec\eta-\vec\eta',x-x') = r^3 \delta^{(3)}(\vec\eta-\vec\eta')\delta^{(3)}(x-x'), \label{eq:GDequation}
\ee
so that the solution to $D X(\vec\eta) = Y(\vec\eta)$ can be written as 
\be
X(\vec\eta,\vec x) = \int \frac{d^3\vec\eta'}{4 r'^3}d^3{\vec x} \,  G_D(\vec\eta,\vec\eta',\vec x- \vec x') Y(\vec\eta',\vec x').
\ee
The operator $G_D$ is both spatially non-local (since $D$ is quadratic in spatial derivatives, $G_D$ is somewhat analogous to $1/{\grad^2}$) and non-local in $\vec\eta$-space.  
It will therefore be useful to write
\be
G_D(\vec\eta,\vec\eta',\vec x,\vec x') = \int d^3{\bf k} \tilde G_D(\vec\eta,\vec\eta',{\bf k}) e^{i{\bf k}.({\bf x - x'})},
\ee
so that $\tilde G_D$ satisfies
\be
r (-i{\bf k}.\grad_\eta+\kappa) r^3 (-i{\bf k}.\grad_\eta+\kappa) \tfrac{1}{r^2}\tilde G_D(\vec\eta,\vec\eta',{\bf k}) = \bar\delta(\vec\eta-\vec\eta',x-x').\label{KdefG}
\ee
This is particularly easy to invert in a $\vec {\bf k}$-centric cylindrical coordinate system $(z,\eta_\perp,\phi)$ where $z = \vec\eta.\hat {\bf k}$, $\eta_\perp = |\vec\eta\cross\hat{\bf k}|$ and $\phi$ is an angle in the plane transverse to ${\bf k}$.  In these coordinates, $r=\sqrt{z^2 +\eta_\perp^2}$ and $\deltabar = -i |{\bf k}| \partial_z+\mu$.  The solution to \eqref{KdefG} is naturally expressed as 
\bea
\tilde G_D(\vec\eta,\vec\eta',{\bf k}) =  -\frac{4}{ |{\bf k}|^2 } e^{i (z'-z) \kappa/|{\bf k}|} \delta_\perp^{(2)}(\vec\eta-\vec\eta')  K(z,z',\eta_\perp)\label{GDsol}
\eea
where $\delta_\perp^{(2)} = \delta(\eta_\perp^2-{\eta'_\perp}^2) \delta(\phi-\phi')$ is a Delta-function in the transverse plane and $K$ is a Green's function for the ordinary differential equation 
\be
r \partial_z r^3 \partial_z \tfrac{1}{r^2} K(z,z',\eta_\perp) = \delta(z-z'). \label{zGreens}
\ee

This second-order ordinary differential equation has a unique solution that is appropriately bounded at $z \rightarrow \pm \infty$:  
\bea
K(z,z') = K(z',z) & =  &r r' \left[\frac{(r'+s z')(r - s z)}{|\eta_\perp|^2}\right]_{s=\sign(z-z')}\label{zGreenSol} \\
& = & r r' \left[ \frac{r'+z'}{r+z} \right]^s\\
& = & r r' \left[ \frac{\eta'.k_\flat}{\eta.k_\flat}\right]_{k_{\flat} = (s |{\bf k}|,\vec{\bf k})}
\eea
where $r'=\sqrt{\eta_\perp^2+z'^2}$ and $\eta \equiv (r,\vec\eta)$.  As expected, the solution is non-local both in position-space and in $\vec\eta$-space.
The kernel $K(z,z')$ has several useful properties:
 \bea
 r \partial_z \frac{1}{rr'} K(z,z')  & = & \sign(z'-z) \frac{K(z,z')}{rr'} = - r' \partial_z' \frac{K(z,z')}{r r'} \\
 \hspace{0.2in}\nonumber\\
 \partial_z \partial_z' \frac{K(z,z') }{(rr')^2} & = & \frac{\delta(z-z')}{r'^3} + \frac{2 \eta_\perp^2}{r^3 r'^3}.
 \eea
These in turn imply
\be
r \deltabar \frac{G_D(\eta,\eta',x-x')}{rr'} = r' \deltabar' \frac{G_D(\eta,\eta',x-x')}{rr'} \label{appeq:rdr}
\ee
and
\be
r^3r'^3\deltabar \deltabar' \frac{G_D(\eta,\eta',x-x')}{(rr')^2} = 
\bar\delta_{\deltabar}(\vec\eta,\vec\eta',x-x') - \bar\delta(\vec\eta,\vec\eta',x-x')\label{appeq:ddp}
\ee
where
\bea
\bar\delta_{\deltabar}(\vec\eta,\vec\eta',x-x') & \equiv & \int d^3k e^{i{\bf k}.{\bf x}} \tilde\delta_{\deltabar}(\vec\eta,\vec\eta',{\bf k}) \\
\tilde\delta_{\deltabar}(\vec\eta,\vec\eta',{\bf k}) & \equiv&  2 \eta_\perp^2 \delta^{(2)}_\perp(\vec\eta-\vec\eta') e^{i (z'-z) \kappa/|{\bf k}|}.\label{projector}
\eea
The object $\bar\delta_{\deltabar}$ can be interpreted as a projector onto the space of functions $f$ satisfying $\deltabar f = 0$, and will play an important role as the Poission bracket of canonical fields in the $\deltabar A = 0$ gauge.  

\section{Dirac Brackets}\label{app:dirac}
Our final Hamiltonian system has two second-class constraints for every $(\vec\eta,\vec x)$:
\be
\phi_2 =  - (J_B + r\deltabar \Pi_A - r^2 \deltabar^2 A) \quad \chi= r \deltabar A.
\ee
We have shown in the main text that $[\phi_2,\phi_2] = [\chi,\chi]=0$ at equal times. 
But their non-zero Poisson bracket with each other:
\bea
\{\phi_2(\eta,x) , \chi(\eta',x')\}
& = & \int \f{d^3\eta''}{4 r''^3} (r \deltabar \bar\delta_{\eta,\eta''}) (r' \deltabar' \bar\delta_{\eta'\eta''}) \\ 
& = & r\deltabar r'\deltabar' \bar\delta_{\eta\eta'} = r \deltabar r^3 \deltabar r^{-2} \bar\delta_{\eta\eta'} = D \bar\delta_{\eta\eta'}
\eea
means that we must use Dirac rather than Poisson brackets for our quantization.

In particular, having already defined a Green's function for $D$ such that $D G_D(\eta,\eta',x-x') = \bar\delta(\eta,\eta')\delta^{(3)}(x-x')$, we can calculate Dirac brackets as 
\be
\{A(\eta,x),B(\eta',x')\}_{D} = \{A,B'\}_{P} + \{A,\phi_2''\}_{P} G_D(\eta'',\eta''') \{\chi''',B'\}_{P} - 
 \{A,\chi''\}_{P} G_D(\eta'',\eta''') \{\phi_2''',B'\}_{P},
\ee
where integrals $d^3\eta/r^3 d^3 x $ for both $\eta''$ and $\eta'''$ are implicit, we have suppressed the $x$-dependence of $G_D$ for brevity, and we use primes on the operator as shorthand for evaluation at the similarly primed $\eta$ and $x$.  

Writing $\{A,\phi_2''\} = {\cal O}_{A,\phi_2} \bar\delta(\eta,\eta'') \delta^{(3)}(x-x')$ where $O$ depends on $\eta$ but not on $\eta''$, and similarly for other commutators, lets us express this as 
\be
\{A(\eta,x),B(\eta',x')\}_{D} = \{A,B'\}_{P} - 
\left( {\cal O}_{\{A,\phi_2\}} {\cal O}'_{\{B',\chi\}} - {\cal O}_{\{A,\chi\}} {\cal O}'_{\{B',\phi_2\}} \right) G_D(\vec\eta,\vec\eta',\vec x - \vec x').
\ee
Using
\begin{align}
{\cal O}_{\{A,\phi_2\}} & = r^3 \deltabar r^{-2} & \qquad {\cal O}_{\{A,\chi\}} & = 0 \\
{\cal O}_{\{\Pi,\phi_2\}} & = r^3 \deltabar^2 r^{-1} & \qquad {\cal O}_{\{\Pi,\chi\}} & = -r^3 \deltabar r^{-2}
\end{align}
we can easily verify (using \eqref{appeq:rdr} in the last line) that the fields have vanishing Dirac brackets with themselves:
\bea
\{A,A'\}_D & = & 0 \\
\{\Pi,\Pi'\}_D & = & \left(-{\cal O}_{\Pi,\phi_2}{\cal O}'_{\Pi,\chi} +  {\cal O}_{\Pi,\chi}{\cal O}'_{\Pi,\phi_2}\right) G_D(\vec\eta,\vec\eta',x-x') \nonumber\\
 & = & (r^3 r'^3 \deltabar \deltabar' \tfrac{1}{rr'}) (- r \deltabar + r' \deltabar) \frac{G_D}{rr'} =0
\eea
and (using \eqref{appeq:ddp}) that the Poisson bracket of $A$ with $\Pi$ is given by
\bea
\{A,\Pi'\}_D & = & \bar\delta(\vec\eta,\vec\eta',x-x') - {\cal O}_{A,\phi_2}{\cal O}'_{\Pi,\chi} G_D(\vec\eta,\vec\eta',x-x') \\
 & = & \bar\delta(\vec\eta,\vec\eta',x-x') + r^3 r'^3 \deltabar \deltabar' \frac{G_D(\vec\eta,\vec\eta',x-x')}{r^2r'^2} \\
& = & \bar\delta_{\deltabar}(\vec\eta,\vec\eta',x-x') \\ 
\eea
where $\bar\delta_{\deltabar}$ is the projector onto the $\bar\delta f = 0$ space defined by \eqref{projector}, as is familiar from other gauge theories.  

\section{Canonical \Poincare Generators}\label{app:poincare}
In this section, we first construct the canonical \Poincare generators for the free CSP theory without gauge fixing and expand them in the $\eta^0=0$ gauge.  We then consider the gauge-fixed form in $\deltabar A = 0$ gauge and their commutators with the canonical $\deltabar A =0$ gauge-fixed fields and with each other.

\subsection{Definition and Derivation of the Generators in Any Gauge}
Canonical \Poincare generators can always be obtained from the Belinfante stress-energy tensor
\be
\Theta^{\mu\nu}(x) \equiv g^{\mu\nu}{\cal L}(x) - \int d^4\eta \frac{\delta {\cal L}}{\delta (\partial_\mu \psi)} \left(\partial^{\nu} \psi\right) 
+ \frac{i}{2} \partial_{\kappa} \left(A^{\kappa \mu\nu} - A^{\mu\kappa\nu}-A^{\nu\kappa\mu} \right)
\ee
where 
\be
A^{\kappa\mu\nu} \equiv \frac{i}{2} \int d^4 \eta \frac{\delta{ \cal L}}{\delta (\partial_{\kappa} \psi)} {\cal J}^{\mu\nu} \psi(\eta,x) \quad \mbox{ and } \quad 
{\cal J}^{\mu\nu} \equiv -i \eta^{[\mu}\partial_{\eta}^{\nu]}.
\ee
and ${\cal J}^{\mu\nu}$ represents an infinitesimal Lorentz generator on the $\eta$-space.
Here and throughout this section, we use $\eta^{[\mu}\partial_{\eta}^{\nu]} = \eta^{\mu}\partial_{\eta}^{\nu} - \eta^{\nu}\partial_{\eta}^{\mu}$. 

Because $\Theta^{\mu\nu}$ is conserved and symmetric, the three-tensor 
\be
{\cal M}^{\kappa\mu\nu} \equiv x^{\mu} \Theta^{\kappa\nu} - x^\nu\Theta^{\kappa\mu} 
\ee
is also conserved.  Its conserved charges 
\bea
J^{\mu\nu} & \equiv & \int d^3 x {\cal M}^{0\mu\nu} = \int d^3x  \left( x^{[\mu} T^{0\nu]} - 2 A^{0\mu\nu} \right)\\
& = & \int d^3 x \; x^{[\mu} g^{0\nu]}{\cal L}(x) - \int d^3x d^4\eta \frac{\delta {\cal L}}{\delta \dot \psi} 
\left( x^{[\mu}  \partial^{\nu]} + \eta^{[\mu}\partial_{\eta}^{\nu]}\right) \psi(\eta,x).
\eea
are the generators of homogeneous Lorentz transformations on fields.  As usual, momentum generators are given by 
\be
P^{i} = \int d^3x  T^{0i} = - \int d^3x d^4\eta \frac{\delta {\cal L}}{\delta \dot \psi} \partial^{i} \psi.
\ee
We adopt the raising and lowering convention that indices on ``$\partial^i$'' are always raised and lowered by $g^{ij}=-\delta^{ij}$ while indices on $\nabla$ are raised and lowered with $\delta^{ij}$, so that 
\be
\partial^{i} = g^{ij} \partial_j = - \partial_i = -\nabla_{i} = -\nabla^i.
\ee 

Using the above definitions and the decomposition $\psi(\eta,x) = A(\vec \eta,x) +\frac{\eta^0}{r} B(\vec\eta,x)+\dots$, it is easy to calculate the generators in terms of the fields $A$ and $B$ and their time derivatives.  As expected, all dependence on $\dot B$ cancels out.  

\bea
P^i & = & - \int d^3x d^4\eta \delta'(\eta^2) \dot \psi \partial^i  \psi +\frac{1}{2} \delta(\eta^2) \partial_\eta.\partial_x \psi \partial_\eta^0 \partial_x^i \psi \\
 & = & \int d^3 x d^4 \eta \tfrac{1}{2} \delta(\eta^2) \left[ \tfrac{1}{r^2} \dot A \nabla^i A 
+ \left(\tfrac{1}{r} \deltabar A +\deltabar \tfrac{1}{r} B\right) \nabla^i B \right] \\
 & = & \int d^3 x \frac{d^3\vec\eta}{4|{\vec\eta}|^3} \; \Pi_A \nabla^i A,
\eea
where the second line follows by substitution of the $A-B$ decomposition for $\psi$ and canceling the $\dot B$ terms, and the third line is obtained by substituting 
\bea
\frac{d^3\vec\eta}{4|{\vec\eta}|^3} = \int d\eta^0 d^4\eta \tfrac{1}{2} \delta(\eta^2) r^{-3}\\
\dot A = \Pi_A + r^3 \deltabar \tfrac{1}{r^2} B
\eea
and integrating by parts so that all $B$ terms cancel. 
% see NT notebook p 18

Proceeding similarly for the rotations,
\bea
J^{ij} &= &  - \int d^3x d^4\eta \delta'(\eta^2) \dot \psi \left( x^{[i} \partial^{j]} + \eta^{[i} \partial_\eta^{j]} \right) \psi +\frac{1}{2} \delta(\eta^2) \partial_\eta.\partial_x \psi \partial_\eta^0 \left( x^{[i} \partial^{j]} + \eta^{[i} \partial_\eta^{j]} \right) \psi  \\
 & = & \int d^3x \frac{d^3\vec\eta}{4} \left[ \tfrac{1}{r^3} \dot A {\cal R}^{ij} A 
+ \deltabar A {\cal R}^{ij} \tfrac{B}{r^2} + \deltabar \tfrac{B}{r} R^{ij} \tfrac{B}{r}\right],
\eea
where
\be
R^{ij} \equiv \left( x^{[i} \nabla^{j]} + \eta^{[i} \nabla_\eta^{j]} \right).
\ee
Using $[\deltabar,R^{ij}] = 0$ and $\nabla^{[j} x^{i]} = x^{[i} \nabla^{j]}$, we can again integrate by parts to simplify the above expression.  As one explicit example, 
\be
\deltabar \tfrac{B}{r} R^{ij} \tfrac{B}{r} \sim \tfrac{B}{r} R^{ij} \deltabar \tfrac{B}{r} \sim - R^{ij} \tfrac{B}{r} \deltabar \tfrac{B}{r}
\ee
where $\sim$ denotes equality up to total derivatives, which implies $\int \deltabar \tfrac{B}{r} R^{ij} \tfrac{B}{r}  = 0$, and similarly cancel the $B$ term from $\dot A$ against $\deltabar A {\cal R}^{ij} \tfrac{B}{r^2}$, leaving
\be
J^{ij}  = \int d^3x \frac{d^3\vec\eta}{4|{\vec\eta}|^3} \left[ \tfrac{1}{r^2} \Pi_A {\cal R}^{ij} A  \right].
\ee

Finally, we can perform a similar procedure for boosts:
\bea
K^{0i} & = & x^0 P^i - \int d^3x \;  x^i {\cal H}(x) +\nonumber \\
& & \qquad \int d^3x \frac{d^3\vec\eta}{4} \left[ \tfrac{1}{r^2} B \nabla_\eta^i (\Pi_A + r^3 \deltabar \tfrac{1}{r^2} B) + (\tfrac{1}{r} \deltabar A + \deltabar \tfrac{1}{r} B) (\nabla_\eta^i A + 2 r \nabla_\eta^i \tfrac{1}{r} B) \right]\\
& =& x^0 P^i - \int d^3x \;  x^i {\cal H}(x) + \nonumber \\
&& \qquad \int d^3x \frac{d^3\vec\eta}{4|{\vec\eta}|^3} \left[r B \nabla_\eta^i \Pi_A + \tfrac{r\eta^i}{2} B \deltabar \tfrac{B}{r} + r^3 \deltabar A \nabla_\eta^i \tfrac{1}{r} B + r^2 \deltabar A \nabla_\eta^i A \right]. \label{bestK}
\eea
For our purposes here, it will be sufficient to consider the special case (for $\deltabar A =0$ gauge in vacuum, in which $\deltabar A = B = 0$
\bea
K^{0i} &\rightarrow& x^0 P^i - \int d^3x \; x^i {\cal H}(x).\\
& = &   x^0 P^i - \int d^3x \frac{d^3\eta}{4r^3}\; \left[\Pi_A x^i \Pi_A + (\grad A) x^i (\grad A)\right].
\eea

\subsection{The $\deltabar=0$ Gauge \Poincare Algebra}
In $\deltabar A = 0$ gauge, and using the commutation relation
\be
[A(\vec\eta,{\bf x} ),\Pi_\perp(\vec\eta',{\bf x'})] = \delta_{\perp,\rho}({\mathbf x-x'},\vec\eta,\vec\eta'),
\ee
we find field commutators
\bea
&&[{\bf J}^{i},A(\vec \eta,{\bf x})] = -i ({\bf x}\cross \grad_x + \vec \eta \cross \grad_\eta)^i A(\vec\eta,{\bf x}) \\
&&[{\bf J}^{i},\Pi_\perp(\vec \eta,{\bf x})] = -i ({\bf x}\cross \grad_x + \vec \eta \cross \grad_\eta)^i \Pi_{\perp}(\vec\eta,{\bf x}) \\
&&[{\bf K}^i,A(\vec \eta,{\bf x})] = -i x^0 \nabla_i A(\vec \eta,{\bf x})  + i \left(x^i -  \f{\eta.\grad_x \grad_\eta^i}{\grad^2_x}\right) \Pi_{\perp}(\vec\eta,{\bf x}) \label{KA}\\
&&[{\bf K}^i,\Pi_\perp(\vec \eta,{\bf x})] = -i x^0 \nabla_i \Pi_\perp(\vec \eta,{\bf x})  + i \left(x^i \grad^2_x -  \eta.\grad_x \grad_\eta^i\right) A(\vec\eta,{\bf x}). \label{KPi}
\eea
The $J$ commutators are easily verified because the rotation commutes with $\deltabar$.  In the second case, the transformation can be obtained by adding total derivatives to the definition of ${\bf K^i}$: in particular, $f \frac{\eta.\grad_x}{\grad_x^2} \grad_\eta^i f$ integrates to zero for any $f$ such that $\deltabar f =0$, and can be freely added to each $x^i$ term in $K^i$.  This is useful because $[\deltabar, x^i -  \frac{\eta.\grad_x}{\grad_x^2} \grad_\eta^i] =  0$, and so this modified boost operator is ``transverse'' and the   $\delta_{\perp,\rho}$ acts on $\left(x^i -  \frac{\eta.\grad_x}{\grad_x^2} \grad_\eta^i\right) f$ as an ordinary $\delta$-function.  Using this trick, it is very easy to derive \eqref{KA} and \eqref{KPi}.  
%%%%%%%%%%%%%%%%%
\bibliography{CSP_GaugeTheory}

\begin{thebibliography}{35}%
\makeatletter
\providecommand \@ifxundefined [1]{%
 \@ifx{#1\undefined}
}%
\providecommand \@ifnum [1]{%
 \ifnum #1\expandafter \@firstoftwo
 \else \expandafter \@secondoftwo
 \fi
}%
\providecommand \@ifx [1]{%
 \ifx #1\expandafter \@firstoftwo
 \else \expandafter \@secondoftwo
 \fi
}%
\providecommand \natexlab [1]{#1}%
\providecommand \enquote  [1]{``#1''}%
\providecommand \bibnamefont  [1]{#1}%
\providecommand \bibfnamefont [1]{#1}%
\providecommand \citenamefont [1]{#1}%
\providecommand \href@noop [0]{\@secondoftwo}%
\providecommand \href [0]{\begingroup \@sanitize@url \@href}%
\providecommand \@href[1]{\@@startlink{#1}\@@href}%
\providecommand \@@href[1]{\endgroup#1\@@endlink}%
\providecommand \@sanitize@url [0]{\catcode `\\12\catcode `\$12\catcode
  `\&12\catcode `\#12\catcode `\^12\catcode `\_12\catcode `\%12\relax}%
\providecommand \@@startlink[1]{}%
\providecommand \@@endlink[0]{}%
\providecommand \url  [0]{\begingroup\@sanitize@url \@url }%
\providecommand \@url [1]{\endgroup\@href {#1}{\urlprefix }}%
\providecommand \urlprefix  [0]{URL }%
\providecommand \Eprint [0]{\href }%
\providecommand \doibase [0]{http://dx.doi.org/}%
\providecommand \selectlanguage [0]{\@gobble}%
\providecommand \bibinfo  [0]{\@secondoftwo}%
\providecommand \bibfield  [0]{\@secondoftwo}%
\providecommand \translation [1]{[#1]}%
\providecommand \BibitemOpen [0]{}%
\providecommand \bibitemStop [0]{}%
\providecommand \bibitemNoStop [0]{.\EOS\space}%
\providecommand \EOS [0]{\spacefactor3000\relax}%
\providecommand \BibitemShut  [1]{\csname bibitem#1\endcsname}%
\let\auto@bib@innerbib\@empty
%</preamble>
\bibitem [{\citenamefont {Wigner}(1939)}]{Wigner:1939cj}%
  \BibitemOpen
  \bibfield  {author} {\bibinfo {author} {\bibfnamefont {Eugene~P.}\
  \bibnamefont {Wigner}},\ }\bibfield  {title} {\enquote {\bibinfo {title} {{On
  Unitary Representations of the Inhomogeneous Lorentz Group}},}\ }\href@noop
  {} {\bibfield  {journal} {\bibinfo  {journal} {Annals Math.}\ }\textbf
  {\bibinfo {volume} {40}},\ \bibinfo {pages} {149--204} (\bibinfo {year}
  {1939})}\BibitemShut {NoStop}%
%%CITATION = ANMAA,40,149;%%
\bibitem [{\citenamefont {Schuster}\ and\ \citenamefont
  {Toro}(2013{\natexlab{a}})}]{Schuster:2013pxj}%
  \BibitemOpen
  \bibfield  {author} {\bibinfo {author} {\bibfnamefont {Philip}\ \bibnamefont
  {Schuster}}\ and\ \bibinfo {author} {\bibfnamefont {Natalia}\ \bibnamefont
  {Toro}},\ }\bibfield  {title} {\enquote {\bibinfo {title} {{On the Theory of
  Continuous-Spin Particles: Wavefunctions and Soft-Factor Scattering
  Amplitudes}},}\ }\href@noop {} {\  (\bibinfo {year} {2013}{\natexlab{a}})},\
  \Eprint {http://arxiv.org/abs/1302.1198} {arXiv:1302.1198 [hep-th]}
  \BibitemShut {NoStop}%
%%CITATION = ARXIV:1302.1198;%%
\bibitem [{\citenamefont {Schuster}\ and\ \citenamefont
  {Toro}(2013{\natexlab{b}})}]{Schuster:2}%
  \BibitemOpen
  \bibfield  {author} {\bibinfo {author} {\bibfnamefont {Philip}\ \bibnamefont
  {Schuster}}\ and\ \bibinfo {author} {\bibfnamefont {Natalia}\ \bibnamefont
  {Toro}},\ }\bibfield  {title} {\enquote {\bibinfo {title} {{On the Theory of
  Continuous-Spin Particles: Helicity Correspondence in Radiation and
  Forces}},}\ }\href@noop {} {\  (\bibinfo {year} {2013}{\natexlab{b}})},\
  \Eprint {http://arxiv.org/abs/1302.1577} {arXiv:1302.1577 [hep-th]}
  \BibitemShut {NoStop}%
%%CITATION = ARXIV:1302.1577;%%
\bibitem [{\citenamefont {Yngvason}(1970)}]{Yngvason:1970fy}%
  \BibitemOpen
  \bibfield  {author} {\bibinfo {author} {\bibfnamefont {J.}~\bibnamefont
  {Yngvason}},\ }\bibfield  {title} {\enquote {\bibinfo {title} {{Zero-mass
  infinite spin representations of the poincare group and quantum field
  theory}},}\ }\href {\doibase 10.1007/BF01649432} {\bibfield  {journal}
  {\bibinfo  {journal} {Commun.Math.Phys.}\ }\textbf {\bibinfo {volume} {18}},\
  \bibinfo {pages} {195--203} (\bibinfo {year} {1970})}\BibitemShut {NoStop}%
%%CITATION = CMPHA,18,195;%%
\bibitem [{\citenamefont {Iverson}\ and\ \citenamefont
  {Mack}(1971)}]{Iverson:1971hq}%
  \BibitemOpen
  \bibfield  {author} {\bibinfo {author} {\bibfnamefont {G.J.}\ \bibnamefont
  {Iverson}}\ and\ \bibinfo {author} {\bibfnamefont {G.}~\bibnamefont {Mack}},\
  }\bibfield  {title} {\enquote {\bibinfo {title} {{Quantum fields and
  interactions of massless particles - the continuous spin case}},}\ }\href
  {\doibase 10.1016/0003-4916(71)90284-3} {\bibfield  {journal} {\bibinfo
  {journal} {Annals Phys.}\ }\textbf {\bibinfo {volume} {64}},\ \bibinfo
  {pages} {211--253} (\bibinfo {year} {1971})}\BibitemShut {NoStop}%
%%CITATION = APNYA,64,211;%%
\bibitem [{\citenamefont {Chakrabarti}(1971)}]{Chakrabarti:1971rz}%
  \BibitemOpen
  \bibfield  {author} {\bibinfo {author} {\bibfnamefont {A.}~\bibnamefont
  {Chakrabarti}},\ }\bibfield  {title} {\enquote {\bibinfo {title} {{Remarks on
  lightlike continuous spin and spacelike representations of the poincare,care
  group}},}\ }\href {\doibase 10.1063/1.1665809} {\bibfield  {journal}
  {\bibinfo  {journal} {J.Math.Phys.}\ }\textbf {\bibinfo {volume} {12}},\
  \bibinfo {pages} {1813--1822} (\bibinfo {year} {1971})}\BibitemShut {NoStop}%
%%CITATION = JMAPA,12,1813;%%
\bibitem [{\citenamefont {Abbott}(1976)}]{Abbott:1976bb}%
  \BibitemOpen
  \bibfield  {author} {\bibinfo {author} {\bibfnamefont {L.F.}\ \bibnamefont
  {Abbott}},\ }\bibfield  {title} {\enquote {\bibinfo {title} {{Massless
  Particles with Continuous Spin Indices}},}\ }\href {\doibase
  10.1103/PhysRevD.13.2291} {\bibfield  {journal} {\bibinfo  {journal}
  {Phys.Rev.}\ }\textbf {\bibinfo {volume} {D13}},\ \bibinfo {pages} {2291}
  (\bibinfo {year} {1976})}\BibitemShut {NoStop}%
%%CITATION = PHRVA,D13,2291;%%
\bibitem [{\citenamefont {Hirata}(1977)}]{Hirata:1977ss}%
  \BibitemOpen
  \bibfield  {author} {\bibinfo {author} {\bibfnamefont {K.}~\bibnamefont
  {Hirata}},\ }\bibfield  {title} {\enquote {\bibinfo {title} {{Quantization of
  Massless Fields with Continuous Spin}},}\ }\href {\doibase
  10.1143/PTP.58.652} {\bibfield  {journal} {\bibinfo  {journal}
  {Prog.Theor.Phys.}\ }\textbf {\bibinfo {volume} {58}},\ \bibinfo {pages}
  {652--666} (\bibinfo {year} {1977})}\BibitemShut {NoStop}%
%%CITATION = PTPKA,58,652;%%
\bibitem [{\citenamefont {Fronsdal}(1978)}]{Fronsdal:1978rb}%
  \BibitemOpen
  \bibfield  {author} {\bibinfo {author} {\bibfnamefont {Christian}\
  \bibnamefont {Fronsdal}},\ }\bibfield  {title} {\enquote {\bibinfo {title}
  {{Massless Fields with Integer Spin}},}\ }\href {\doibase
  10.1103/PhysRevD.18.3624} {\bibfield  {journal} {\bibinfo  {journal}
  {Phys.Rev.}\ }\textbf {\bibinfo {volume} {D18}},\ \bibinfo {pages} {3624}
  (\bibinfo {year} {1978})}\BibitemShut {NoStop}%
%%CITATION = PHRVA,D18,3624;%%
\bibitem [{\citenamefont {Sorokin}(2005)}]{Sorokin:2004ie}%
  \BibitemOpen
  \bibfield  {author} {\bibinfo {author} {\bibfnamefont {Dmitri}\ \bibnamefont
  {Sorokin}},\ }\bibfield  {title} {\enquote {\bibinfo {title} {{Introduction
  to the classical theory of higher spins}},}\ }\href {\doibase
  10.1063/1.1923335} {\bibfield  {journal} {\bibinfo  {journal} {AIP
  Conf.Proc.}\ }\textbf {\bibinfo {volume} {767}},\ \bibinfo {pages} {172--202}
  (\bibinfo {year} {2005})},\ \Eprint {http://arxiv.org/abs/hep-th/0405069}
  {arXiv:hep-th/0405069 [hep-th]} \BibitemShut {NoStop}%
%%CITATION = HEP-TH/0405069;%%
\bibitem [{\citenamefont {Bargmann}(1947)}]{Bargmann:1946me}%
  \BibitemOpen
  \bibfield  {author} {\bibinfo {author} {\bibfnamefont {V.}~\bibnamefont
  {Bargmann}},\ }\bibfield  {title} {\enquote {\bibinfo {title} {{Irreducible
  unitary representations of the Lorentz group}},}\ }\href@noop {} {\bibfield
  {journal} {\bibinfo  {journal} {Annals Math.}\ }\textbf {\bibinfo {volume}
  {48}},\ \bibinfo {pages} {568--640} (\bibinfo {year} {1947})}\BibitemShut
  {NoStop}%
%%CITATION = ANMAA,48,568;%%
\bibitem [{\citenamefont {Bargmann}\ and\ \citenamefont
  {Wigner}(1948)}]{Bargmann:1948ck}%
  \BibitemOpen
  \bibfield  {author} {\bibinfo {author} {\bibfnamefont {V.}~\bibnamefont
  {Bargmann}}\ and\ \bibinfo {author} {\bibfnamefont {Eugene~P.}\ \bibnamefont
  {Wigner}},\ }\bibfield  {title} {\enquote {\bibinfo {title} {{Group
  Theoretical Discussion of Relativistic Wave Equations}},}\ }\href@noop {}
  {\bibfield  {journal} {\bibinfo  {journal} {Proc.Nat.Acad.Sci.}\ }\textbf
  {\bibinfo {volume} {34}},\ \bibinfo {pages} {211} (\bibinfo {year}
  {1948})}\BibitemShut {NoStop}%
%%CITATION = PNASA,34,211;%%
\bibitem [{\citenamefont {Wigner}(1963)}]{Wigner:1963}%
  \BibitemOpen
  \bibfield  {author} {\bibinfo {author} {\bibfnamefont {Eugene~P.}\
  \bibnamefont {Wigner}},\ }\bibfield  {title} {\enquote {\bibinfo {title}
  {{Invariant Quantum Mechanical Equations of Motion}},}\ }\href@noop {}
  {\bibfield  {journal} {\bibinfo  {journal} {Theoretical Physics,
  International Atomic Energy Agency, Vienna}\ } (\bibinfo {year}
  {1963})}\BibitemShut {NoStop}%
%%CITATION
\bibitem [{Note1()}]{Note1}%
  \BibitemOpen
  \bibinfo {note} {Having classified all covariant wave equations for CSPs
  \protect \emph {without} gauge redundancy \cite {Schuster:2013pxj}, it is now
  clear that none of these can be used to uniquely relate single-particle
  states to canonically quantized, local fields.}\BibitemShut {Stop}%
\bibitem [{\citenamefont {Bekaert}\ and\ \citenamefont
  {Mourad}(2006)}]{Bekaert:2005in}%
  \BibitemOpen
  \bibfield  {author} {\bibinfo {author} {\bibfnamefont {X.}~\bibnamefont
  {Bekaert}}\ and\ \bibinfo {author} {\bibfnamefont {J.}~\bibnamefont
  {Mourad}},\ }\bibfield  {title} {\enquote {\bibinfo {title} {{The Continuous
  spin limit of higher spin field equations}},}\ }\href {\doibase
  10.1088/1126-6708/2006/01/115} {\bibfield  {journal} {\bibinfo  {journal}
  {JHEP}\ }\textbf {\bibinfo {volume} {0601}},\ \bibinfo {pages} {115}
  (\bibinfo {year} {2006})},\ \Eprint {http://arxiv.org/abs/hep-th/0509092}
  {arXiv:hep-th/0509092 [hep-th]} \BibitemShut {NoStop}%
%%CITATION = HEP-TH/0509092;%%
\bibitem [{Note2()}]{Note2}%
  \BibitemOpen
  \bibinfo {note} {The Fronsdal-like equation of motion of \cite
  {Bekaert:2005in} can be obtained by localizing \protect \textup {\hbox
  {\mathsurround \z@ \protect \normalfont (\ignorespaces \ref
  {eq:BaseAction}\unskip \@@italiccorr )}} on $\delta '(\eta ^2+1)$ in our
  metric conventions and similarly modifying \protect \textup {\hbox
  {\mathsurround \z@ \protect \normalfont (\ignorespaces \ref
  {gaugePsiA}\unskip \@@italiccorr )}}. This localization obstructs the
  restriction of \protect \textup {\hbox {\mathsurround \z@ \protect
  \normalfont (\ignorespaces \ref {eq:BaseAction}\unskip \@@italiccorr )}} to
  positive $\eta ^0$ and the simple decomposition of $\psi $ in Section \ref
  {dynamical}, but alternate decompositions and subsequent quantizations may be
  possible. Like our equations of motion, the Fronsdal-like equation of \cite
  {Bekaert:2005in} propagates CSPs of all $\rho $ unless further non-local
  restrictions are imposed. An additional feature of \protect \textup {\hbox
  {\mathsurround \z@ \protect \normalfont (\ignorespaces \ref {eomPsi1}\unskip
  \@@italiccorr )}} and \protect \textup {\hbox {\mathsurround \z@ \protect
  \normalfont (\ignorespaces \ref {eomPsi2}\unskip \@@italiccorr )}} relative
  to those of \cite {Bekaert:2005in}, which may or may not be significant, is
  that \protect \textup {\hbox {\mathsurround \z@ \protect \normalfont
  (\ignorespaces \ref {eomPsi1}\unskip \@@italiccorr )}} and \protect \textup
  {\hbox {\mathsurround \z@ \protect \normalfont (\ignorespaces \ref
  {eomPsi2}\unskip \@@italiccorr )}} precisely recover the Fronsdal equations
  in the $\kappa \rightarrow 0$ limit.}\BibitemShut {Stop}%
\bibitem [{\citenamefont {Bouatta}\ \emph {et~al.}(2004)\citenamefont
  {Bouatta}, \citenamefont {Compere},\ and\ \citenamefont
  {Sagnotti}}]{Bouatta:2004kk}%
  \BibitemOpen
  \bibfield  {author} {\bibinfo {author} {\bibfnamefont {N.}~\bibnamefont
  {Bouatta}}, \bibinfo {author} {\bibfnamefont {G.}~\bibnamefont {Compere}}, \
  and\ \bibinfo {author} {\bibfnamefont {A.}~\bibnamefont {Sagnotti}},\
  }\bibfield  {title} {\enquote {\bibinfo {title} {{An Introduction to free
  higher-spin fields}},}\ }\href@noop {} {\  (\bibinfo {year} {2004})},\
  \Eprint {http://arxiv.org/abs/hep-th/0409068} {arXiv:hep-th/0409068 [hep-th]}
  \BibitemShut {NoStop}%
%%CITATION = HEP-TH/0409068;%%
\bibitem [{\citenamefont {Bekaert}\ \emph {et~al.}(2005)\citenamefont
  {Bekaert}, \citenamefont {Cnockaert}, \citenamefont {Iazeolla},\ and\
  \citenamefont {Vasiliev}}]{Bekaert:2005vh}%
  \BibitemOpen
  \bibfield  {author} {\bibinfo {author} {\bibfnamefont {X.}~\bibnamefont
  {Bekaert}}, \bibinfo {author} {\bibfnamefont {S.}~\bibnamefont {Cnockaert}},
  \bibinfo {author} {\bibfnamefont {Carlo}\ \bibnamefont {Iazeolla}}, \ and\
  \bibinfo {author} {\bibfnamefont {M.A.}\ \bibnamefont {Vasiliev}},\
  }\bibfield  {title} {\enquote {\bibinfo {title} {{Nonlinear higher spin
  theories in various dimensions}},}\ }\href@noop {} {\  (\bibinfo {year}
  {2005})},\ \Eprint {http://arxiv.org/abs/hep-th/0503128}
  {arXiv:hep-th/0503128 [hep-th]} \BibitemShut {NoStop}%
%%CITATION = HEP-TH/0503128;%%
\bibitem [{\citenamefont {Fotopoulos}\ and\ \citenamefont
  {Tsulaia}(2009)}]{Fotopoulos:2008ka}%
  \BibitemOpen
  \bibfield  {author} {\bibinfo {author} {\bibfnamefont {Angelos}\ \bibnamefont
  {Fotopoulos}}\ and\ \bibinfo {author} {\bibfnamefont {Mirian}\ \bibnamefont
  {Tsulaia}},\ }\bibfield  {title} {\enquote {\bibinfo {title} {{Gauge
  Invariant Lagrangians for Free and Interacting Higher Spin Fields. A Review
  of the BRST formulation}},}\ }\href {\doibase 10.1142/S0217751X09043134}
  {\bibfield  {journal} {\bibinfo  {journal} {Int.J.Mod.Phys.}\ }\textbf
  {\bibinfo {volume} {A24}},\ \bibinfo {pages} {1--60} (\bibinfo {year}
  {2009})},\ \Eprint {http://arxiv.org/abs/0805.1346} {arXiv:0805.1346
  [hep-th]} \BibitemShut {NoStop}%
%%CITATION = ARXIV:0805.1346;%%
\bibitem [{\citenamefont {Benincasa}\ and\ \citenamefont
  {Conde}(2012)}]{Benincasa:2011pg}%
  \BibitemOpen
  \bibfield  {author} {\bibinfo {author} {\bibfnamefont {Paolo}\ \bibnamefont
  {Benincasa}}\ and\ \bibinfo {author} {\bibfnamefont {Eduardo}\ \bibnamefont
  {Conde}},\ }\bibfield  {title} {\enquote {\bibinfo {title} {{Exploring the
  S-Matrix of Massless Particles}},}\ }\href {\doibase
  10.1103/PhysRevD.86.025007} {\bibfield  {journal} {\bibinfo  {journal}
  {Phys.Rev.}\ }\textbf {\bibinfo {volume} {D86}},\ \bibinfo {pages} {025007}
  (\bibinfo {year} {2012})},\ \Eprint {http://arxiv.org/abs/1108.3078}
  {arXiv:1108.3078 [hep-th]} \BibitemShut {NoStop}%
%%CITATION = ARXIV:1108.3078;%%
\bibitem [{\citenamefont {Schuster}\ and\ \citenamefont
  {Toro}(2013{\natexlab{c}})}]{ST:inprep}%
  \BibitemOpen
  \bibfield  {author} {\bibinfo {author} {\bibfnamefont {Philip}\ \bibnamefont
  {Schuster}}\ and\ \bibinfo {author} {\bibfnamefont {Natalia}\ \bibnamefont
  {Toro}},\ }\href@noop {} {\bibfield  {journal} {\bibinfo  {journal} {in
  preparation}\ } (\bibinfo {year} {2013}{\natexlab{c}})}\BibitemShut {NoStop}%
%%CITATION
\bibitem [{\citenamefont {Weinberg}(1965)}]{Weinberg:1965rz}%
  \BibitemOpen
  \bibfield  {author} {\bibinfo {author} {\bibfnamefont {Steven}\ \bibnamefont
  {Weinberg}},\ }\bibfield  {title} {\enquote {\bibinfo {title} {{Photons and
  gravitons in perturbation theory: Derivation of Maxwell's and Einstein's
  equations}},}\ }\href {\doibase 10.1103/PhysRev.138.B988} {\bibfield
  {journal} {\bibinfo  {journal} {Phys.Rev.}\ }\textbf {\bibinfo {volume}
  {138}},\ \bibinfo {pages} {B988--B1002} (\bibinfo {year} {1965})}\BibitemShut
  {NoStop}%
%%CITATION = PHRVA,138,B988;%%
\bibitem [{\citenamefont {Weinberg}(1964{\natexlab{a}})}]{Weinberg:1964ew}%
  \BibitemOpen
  \bibfield  {author} {\bibinfo {author} {\bibfnamefont {Steven}\ \bibnamefont
  {Weinberg}},\ }\bibfield  {title} {\enquote {\bibinfo {title} {{Photons and
  Gravitons in s Matrix Theory: Derivation of Charge Conservation and Equality
  of Gravitational and Inertial Mass}},}\ }\href {\doibase
  10.1103/PhysRev.135.B1049} {\bibfield  {journal} {\bibinfo  {journal}
  {Phys.Rev.}\ }\textbf {\bibinfo {volume} {135}},\ \bibinfo {pages}
  {B1049--B1056} (\bibinfo {year} {1964}{\natexlab{a}})}\BibitemShut {NoStop}%
%%CITATION = PHRVA,135,B1049;%%
\bibitem [{\citenamefont {Weinberg}(1964{\natexlab{b}})}]{Weinberg:1964ev}%
  \BibitemOpen
  \bibfield  {author} {\bibinfo {author} {\bibfnamefont {Steven}\ \bibnamefont
  {Weinberg}},\ }\bibfield  {title} {\enquote {\bibinfo {title} {{Feynman Rules
  for Any Spin. 2. Massless Particles}},}\ }\href {\doibase
  10.1103/PhysRev.134.B882} {\bibfield  {journal} {\bibinfo  {journal}
  {Phys.Rev.}\ }\textbf {\bibinfo {volume} {134}},\ \bibinfo {pages}
  {B882--B896} (\bibinfo {year} {1964}{\natexlab{b}})}\BibitemShut {NoStop}%
%%CITATION = PHRVA,134,B882;%%
\bibitem [{\citenamefont {Porrati}(2012)}]{Porrati:2012rd}%
  \BibitemOpen
  \bibfield  {author} {\bibinfo {author} {\bibfnamefont {M.}~\bibnamefont
  {Porrati}},\ }\bibfield  {title} {\enquote {\bibinfo {title} {{Old and New No
  Go Theorems on Interacting Massless Particles in Flat Space}},}\ }\href@noop
  {} {\  (\bibinfo {year} {2012})},\ \Eprint {http://arxiv.org/abs/1209.4876}
  {arXiv:1209.4876 [hep-th]} \BibitemShut {NoStop}%
%%CITATION = ARXIV:1209.4876;%%
\bibitem [{\citenamefont {Benincasa}\ and\ \citenamefont
  {Cachazo}(2007)}]{Benincasa:2007xk}%
  \BibitemOpen
  \bibfield  {author} {\bibinfo {author} {\bibfnamefont {Paolo}\ \bibnamefont
  {Benincasa}}\ and\ \bibinfo {author} {\bibfnamefont {Freddy}\ \bibnamefont
  {Cachazo}},\ }\bibfield  {title} {\enquote {\bibinfo {title} {{Consistency
  Conditions on the S-Matrix of Massless Particles}},}\ }\href@noop {} {\
  (\bibinfo {year} {2007})},\ \Eprint {http://arxiv.org/abs/0705.4305}
  {arXiv:0705.4305 [hep-th]} \BibitemShut {NoStop}%
%%CITATION = ARXIV:0705.4305;%%
\bibitem [{\citenamefont {Schuster}\ and\ \citenamefont
  {Toro}(2009)}]{Schuster:2008nh}%
  \BibitemOpen
  \bibfield  {author} {\bibinfo {author} {\bibfnamefont {Philip~C.}\
  \bibnamefont {Schuster}}\ and\ \bibinfo {author} {\bibfnamefont {Natalia}\
  \bibnamefont {Toro}},\ }\bibfield  {title} {\enquote {\bibinfo {title}
  {{Constructing the Tree-Level Yang-Mills S-Matrix Using Complex
  Factorization}},}\ }\href {\doibase 10.1088/1126-6708/2009/06/079} {\bibfield
   {journal} {\bibinfo  {journal} {JHEP}\ }\textbf {\bibinfo {volume} {0906}},\
  \bibinfo {pages} {079} (\bibinfo {year} {2009})},\ \Eprint
  {http://arxiv.org/abs/0811.3207} {arXiv:0811.3207 [hep-th]} \BibitemShut
  {NoStop}%
%%CITATION = ARXIV:0811.3207;%%
\bibitem [{\citenamefont {Heinonen}\ \emph {et~al.}(2012)\citenamefont
  {Heinonen}, \citenamefont {Hill},\ and\ \citenamefont
  {Solon}}]{Heinonen:2012km}%
  \BibitemOpen
  \bibfield  {author} {\bibinfo {author} {\bibfnamefont {Johannes}\
  \bibnamefont {Heinonen}}, \bibinfo {author} {\bibfnamefont {Richard~J.}\
  \bibnamefont {Hill}}, \ and\ \bibinfo {author} {\bibfnamefont {Mikhail~P.}\
  \bibnamefont {Solon}},\ }\bibfield  {title} {\enquote {\bibinfo {title}
  {{Lorentz invariance in heavy particle effective theories}},}\ }\href
  {\doibase 10.1103/PhysRevD.86.094020} {\bibfield  {journal} {\bibinfo
  {journal} {Phys.Rev.}\ }\textbf {\bibinfo {volume} {D86}},\ \bibinfo {pages}
  {094020} (\bibinfo {year} {2012})},\ \Eprint {http://arxiv.org/abs/1208.0601}
  {arXiv:1208.0601 [hep-ph]} \BibitemShut {NoStop}%
%%CITATION = ARXIV:1208.0601;%%
\bibitem [{\citenamefont {Brink}\ \emph {et~al.}(2002)\citenamefont {Brink},
  \citenamefont {Khan}, \citenamefont {Ramond},\ and\ \citenamefont
  {Xiong}}]{Brink:2002zx}%
  \BibitemOpen
  \bibfield  {author} {\bibinfo {author} {\bibfnamefont {Lars}\ \bibnamefont
  {Brink}}, \bibinfo {author} {\bibfnamefont {Abu~M.}\ \bibnamefont {Khan}},
  \bibinfo {author} {\bibfnamefont {Pierre}\ \bibnamefont {Ramond}}, \ and\
  \bibinfo {author} {\bibfnamefont {Xiao-zhen}\ \bibnamefont {Xiong}},\
  }\bibfield  {title} {\enquote {\bibinfo {title} {{Continuous spin
  representations of the Poincare and superPoincare groups}},}\ }\href
  {\doibase 10.1063/1.1518138} {\bibfield  {journal} {\bibinfo  {journal}
  {J.Math.Phys.}\ }\textbf {\bibinfo {volume} {43}},\ \bibinfo {pages} {6279}
  (\bibinfo {year} {2002})},\ \Eprint {http://arxiv.org/abs/hep-th/0205145}
  {arXiv:hep-th/0205145 [hep-th]} \BibitemShut {NoStop}%
%%CITATION = HEP-TH/0205145;%%
\bibitem [{\citenamefont {Bjorken}\ and\ \citenamefont
  {Drell}(1965)}]{Bjorken:1965zz}%
  \BibitemOpen
  \bibfield  {author} {\bibinfo {author} {\bibfnamefont {James~D.}\
  \bibnamefont {Bjorken}}\ and\ \bibinfo {author} {\bibfnamefont {Sidney~D.}\
  \bibnamefont {Drell}},\ }\bibfield  {title} {\enquote {\bibinfo {title}
  {{Relativistic quantum fields}},}\ }\href@noop {} {\  (\bibinfo {year}
  {1965})}\BibitemShut {NoStop}%
%%CITATION = ISBN-0070054940 ETC.;%%
\bibitem [{\citenamefont {Weinberg}(1995)}]{WeinbergQFT}%
  \BibitemOpen
  \bibfield  {author} {\bibinfo {author} {\bibfnamefont {Steven}\ \bibnamefont
  {Weinberg}},\ }\bibfield  {title} {\enquote {\bibinfo {title} {{The Quantum
  Theory of Fields I}},}\ }\href@noop {} {\  (\bibinfo {year}
  {1995})}\BibitemShut {NoStop}%
%%CITATION
\bibitem [{\citenamefont {Dirac}(1964)}]{DiracLectures}%
  \BibitemOpen
  \bibfield  {author} {\bibinfo {author} {\bibfnamefont {P.A.M.}\ \bibnamefont
  {Dirac}},\ }\bibfield  {title} {\enquote {\bibinfo {title} {{Lectures on
  Quantum Mechanics}},}\ }\href@noop {} {\  (\bibinfo {year}
  {1964})}\BibitemShut {NoStop}%
\bibitem [{\citenamefont {Vasiliev}(1988)}]{Vasiliev:1988xc}%
  \BibitemOpen
  \bibfield  {author} {\bibinfo {author} {\bibfnamefont {Mikhail~A.}\
  \bibnamefont {Vasiliev}},\ }\bibfield  {title} {\enquote {\bibinfo {title}
  {{Equations of motion of interacting massless fields of all spins as a free
  differential algebra}},}\ }\href {\doibase 10.1016/0370-2693(88)91179-3}
  {\bibfield  {journal} {\bibinfo  {journal} {Phys.Lett.}\ }\textbf {\bibinfo
  {volume} {B209}},\ \bibinfo {pages} {491--497} (\bibinfo {year}
  {1988})}\BibitemShut {NoStop}%
%%CITATION = PHLTA,B209,491;%%
\bibitem [{\citenamefont {Vasiliev}(1989)}]{Vasiliev:1988sa}%
  \BibitemOpen
  \bibfield  {author} {\bibinfo {author} {\bibfnamefont {Mikhail~A.}\
  \bibnamefont {Vasiliev}},\ }\bibfield  {title} {\enquote {\bibinfo {title}
  {{Consistent equations for interacting massless fields of all spins in the
  first order in curvatures}},}\ }\href {\doibase 10.1016/0003-4916(89)90261-3}
  {\bibfield  {journal} {\bibinfo  {journal} {Annals Phys.}\ }\textbf {\bibinfo
  {volume} {190}},\ \bibinfo {pages} {59--106} (\bibinfo {year}
  {1989})}\BibitemShut {NoStop}%
%%CITATION = APNYA,190,59;%%
\bibitem [{\citenamefont {Vasiliev}(2004)}]{Vasiliev:2004qz}%
  \BibitemOpen
  \bibfield  {author} {\bibinfo {author} {\bibfnamefont {M.A.}\ \bibnamefont
  {Vasiliev}},\ }\bibfield  {title} {\enquote {\bibinfo {title} {{Higher spin
  gauge theories in various dimensions}},}\ }\href {\doibase
  10.1002/prop.200410167} {\bibfield  {journal} {\bibinfo  {journal}
  {Fortsch.Phys.}\ }\textbf {\bibinfo {volume} {52}},\ \bibinfo {pages}
  {702--717} (\bibinfo {year} {2004})},\ \Eprint
  {http://arxiv.org/abs/hep-th/0401177} {arXiv:hep-th/0401177 [hep-th]}
  \BibitemShut {NoStop}%
%%CITATION = HEP-TH/0401177;%%
\end{thebibliography}%

\end{document}